\documentclass[sn-mathphys-num]{sn-jnl}

\usepackage{graphicx}%
\usepackage{multirow}%
\usepackage{amsmath,amssymb,amsfonts}%
\usepackage{amsthm}%
\usepackage{mathrsfs}%
\usepackage[title]{appendix}%
\usepackage{xcolor}%
\usepackage{textcomp}%
\usepackage{manyfoot}%
\usepackage{booktabs}%
\usepackage{algorithm}%
\usepackage{algorithmicx}%
\usepackage{algpseudocode}%
\usepackage{listings}%
\usepackage{gensymb}%
\usepackage{braket}%
\usepackage{siunitx}

\theoremstyle{thmstyleone}%

\theoremstyle{thmstyletwo}%

\theoremstyle{thmstylethree}%

\begin{document}

\title[Article Title]{An experimental platform for studying the heteronuclear Efimov effect with an ultracold mixture of $^6$Li and $^{133}$Cs atoms}

\author*[1]{\fnm{Eleonora} \sur{Lippi}}\email{lippi@physi.uni-heidelberg.de}

\author[1]{\fnm{Manuel} \sur{Gerken}}

\author[1]{\fnm{Stephan} \sur{H\"afner}}

\author[1]{\fnm{Marc} \sur{Repp}}

\author[1]{\fnm{Rico} \sur{Pires}}

\author[1]{\fnm{Michael} \sur{Rautenberg}}

\author[1]{\fnm{Tobias} \sur{Krom}}

\author[1]{\fnm{Eva D.} \sur{Kuhnle}}

\author[1]{\fnm{Binh} \sur{Tran}}

\author[1]{\fnm{Juris} \sur{Ulmanis}}

\author[1]{\fnm{Bing} \sur{Zhu}} \equalcont{current address: HSBC Lab, Guangzhou 510510, China}

\author*[1]{\fnm{Lauriane} \sur{Chomaz}}\email{chomaz@physi.uni-heidelberg.de}

\author*[1]{\fnm{Matthias} \sur{ Weidem\"uller}}\email{weidemueller@uni-heidelberg.de}

\affil[1]{\orgdiv{Physikalisches Institut}, \orgname{Universit\"at Heidelberg}, \orgaddress{\street{Im Neuenheimer Feld, 226}, \city{Heidelberg}, \postcode{69120}, \state{Baden-W\"urttemberg}, \country{Germany}}}

\abstract{We present the experimental apparatus enabling the observation of the heteronuclear Efimov effect in an optically trapped ultracold mixture of $^6$Li-$^{133}$Cs with high-resolution control of the interactions.
A compact double-species Zeeman slower consisting of four interleaving helical coils allows for a fast-switching between two optimized configurations for either Li or Cs and provides an efficient sequential loading into their respective MOTs. By means of a bichromatic optical trapping scheme based on species-selective trapping we prepare mixtures down to \SI{100}{nK} of \num{1e4} Cs atoms and \num{7e3} Li atoms. Highly stable magnetic fields allow high-resolution atom-loss spectroscopy and enable to resolve splitting in the loss feature of a few tens of milligauss. These features allowed for a detailed study of the Efimov effect.}


\maketitle

\section{Introduction}\label{sec1}

The three-body problem in quantum mechanics plays a crucial role in revealing universal phenomena in few-body physics \cite{Greene2010,  Blume2012, Frederico2012, Wang2013}. 
A key example is the Efimov scenario \cite{Efimov1970, Efimov1971, Efimov1973, Efimov1979}, which occurs in a three-body system with pairwise resonant interactions, manifesting in an infinite sequence of bound states known as Efimov states. These energy states exhibit a universal discrete scaling law, for which they are self-similar under rescaling of all the length scales by the discrete factor $\lambda$. This factor depends only on the quantum statistics, the mass ratio, and the number of resonant pairwise interactions of the atoms involved.
This effect, originally predicted by Efimov in nuclear systems \cite{Efimov1970}, such as the $^{12}$C nucleus or tritium ($^{3}$H), was first experimentally observed in ultracold gases \cite{kraemer2006, Ferlaino2011} thanks to the ability of fine-tuning two-body interactions via Feshbach resonances (FRs)  \cite{Chin2010}. After initial studies in homonuclear ultracold systems \cite{kraemer2006, Williams2009, Zaccanti2009}, the Efimov effect was later detected in heteronuclear ultracold systems \cite{Barontini2009, Barontini2010, Bloom2013a, Hu2014a, Pires_2014, Tung2014, Maier2015} and also in molecular beam experiments with the excited helium trimer $^4$He$_3$ \cite{Kunitski2015}. 

\begin{figure}
	\centering
	\includegraphics[scale=0.35]{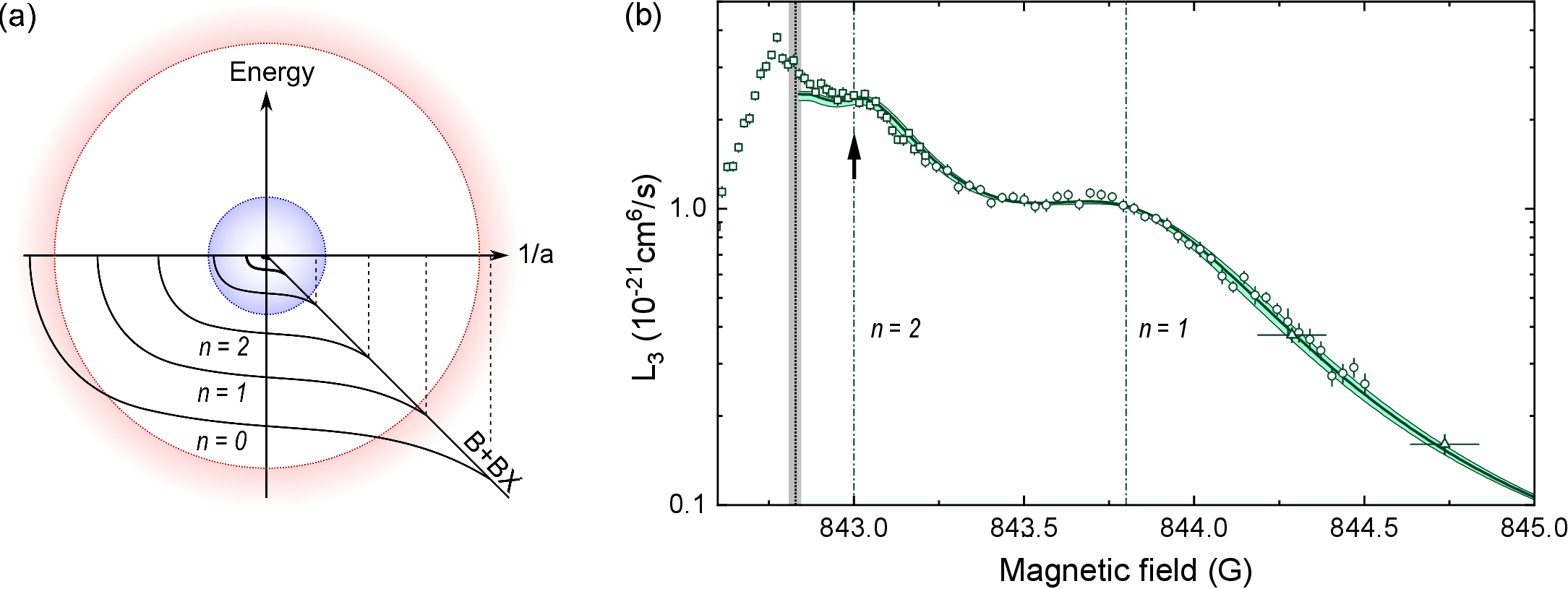}
    \caption{ (a) Cartoon of the heteronulear Efimov scenario for two identical bosons \textit{B} and one distinguishable particle \textit{X}. The deepest Efimov trimers as a function of the interspecies scattering \textit{B-X} are shown. The intraspecies \textit{B-B} scattering length is resonant.  The window of universality (white area) lies in between the regions dominated by temperature (blue shaded area) and short-range two-body interactions (red shaded area), respectively. For $B$ = Cs and $X$ = Li it was possible to access not only the ground ($n=0$) and first excited ($n=1$) Efimov state, but also the second excited ($n=2$) Efimov state. Picture adapted from \citet{Ulmanis2015}. (b) Cs-Cs-Li three-body loss rate $L_3$ at a temperature of \SI{120} {nK}. Experimental data are reproduced using  a fit to the zero-range theory (green dashed line), as discussed in Ref.\cite{Ulmanis_2016a}. The different symbols (squares, circles, triangles) belong to different sets of data. The shaded area around the fits represents the uncertainty on the Feshbach resonance position. The black dashed vertical lines correspond to the positions of the first ($n=1$)  and second ($n=2$)  excited Efimov state as predicted from the theory, respectively at \SI{843.8}{G} and \SI{843.0}{G}. The pole of the Feshbach resonance is indicated by the dotted line and the uncertainty by the gray shade. Data taken from \citet{Ulmanis2015}.}
	\label{fig:3bodyLoss}
\end{figure}

In ultracold heteronuclear gases, the signatures of the Efimov effect are much richer compared to homonuclear systems \cite{Ulmanis2016b}, as a significant  mass ratio results in a denser Efimov spectrum. The Efimov scenario for the heteronuclear case  for two identical bosons \textit{B} and one distinguishable particle \textit{X},  can be represented in the form of an energy diagram as a function of the inverse scattering length, as shown in Fig.~\ref{fig:3bodyLoss}(a). A few of the deepest bound Efimov trimers are illustrated. They connect the dissociation scattering threshold ($E>0$), where three atoms are unbound, to the atom-dimer threshold  ($E<0$, $a>0$), where the system supports a bound dimer \textit{BX} plus a free atom \textit{B}. The universal discrete scaling only holds for scattering lengths that are larger than the length scale characterizing short-range interactions, due to a finite range pairwise potential, and smaller than the De Broglie wavelength, which is set by the temperature of the system. 

The most common approach to probe the Efimov scenario is by using atom-loss spectroscopy. The scattering length is tuned by scanning the magnetic field across a Feshbach resonance and the Efimov resonances emerge as additional modulations on top of the prominent loss feature associated with the Feshbach resonance. A complementary measurement is provided by recording the time evolution of the atom number of each species, which is fitted with coupled rate equations and yields three-body loss rate spectra as illustrated in Fig.~\ref{fig:3bodyLoss} (b).
The extreme mass ratio of $\approx 22$ of the system formed by heavy bosonic $^{133}$Cs and light fermionic $^6$Li leads to a reduction of the Efimov scaling factor, from $\lambda= 22.7$ (for equal-mass bosons) to $\lambda=4.88$, which is roughly a fifth compared to equal-mass systems \cite{DIncao2006}.  This mass ratio in combination to the broad interspecies Feshbach resonances \cite{Repp2013,Tung2013} and the low temperatures reached, allowed the observation of up to three consecutive resonances for the first time \cite{Pires_2014, Tung2014, Ulmanis_2016, Ulmanis_2016a}. 

However, a large mass imbalance poses several experimental challenges for slowing, cooling, and trapping different species together and, therefore, the development of efficient techniques to produce large two-component mass-imbalanced quantum gases deserves particular attention. 
The most commonly used procedure to obtain ultracold gases is to load atoms in a magneto-optical trap (MOT) \cite{Weiner1999, Schunemann1998}, typically followed by supplemental laser cooling schemes \cite{MetcalfStraten1999, Cohen2011} and evaporative cooling \cite{Ketterle1996} in a magnetic or an optical dipole trap (ODT) \cite{Grimm2000}. The first crucial requirement in this scheme is the generation of an atomic source with a high flux of atoms at low velocities for efficient loading into a MOT. One of the most common approaches to achieve this goal is to use a Zeeman slower (ZS) \cite{Phillips1982}. 
Species with large mass difference require very different magnetic field strengths and profiles to efficiently decelerate the individual beams in a ZS and appropriate custom-built solutions are needed.
The second important requirement arises from the combined trapping of different atomic species. Mass imbalance and species dependent polarizability have to be taken into account in order to obtain an efficient transfer of laser cooled atoms into an ODT and maintain a sufficient overlap between the two atomic clouds despite the large relative displacement under the effect of gravity, named as the differential gravitational sag. 
\bigskip

In this paper we present the experimental apparatus enabling the observation of the heteronuclear Efimov effect in an optically trapped ultracold mixture of $^6$Li-$^{133}$Cs with high resolution control of tunable interactions, as reported in \cite{Pires_2014, Ulmanis_2016, Ulmanis_2016a, Haefner_2017}. We focus on the key ingredients of our experimental apparatus concerning slowing of hot atoms, management of dipole trapping and of magnetic fields, respectively.

In Sec.~\ref{sec:slower} we present the design and realization of a double-species Zeeman slower, allowing for fast switching-times between two optimized magnetic field configurations for decelerating either Li or Cs and providing an efficient sequential loading scheme into their respective MOTs. This compact design reduces the cost and complexity of our apparatus and offers sufficient optical access for a double-species experiment.

In Sec.~\ref{sec:mixture} we demonstrate how to successfully realize an optically trapped  $^6$Li-$^{133}$Cs mixture at the temperature of \SI{100}{nK} by means of a combined bichromatic trapping scheme. In order to fulfill the individual requirements for an efficient production of cold samples, we load and evaporatively cool down Li and Cs atoms into spatially separated optical dipole traps and mix them together only at a later stage. 
The method of species-selective trapping \cite{LeBlanc2007} allows us to maintain a partial overlap between the two clouds by exploiting a large difference in the polarizability between the two species.

In Sec.~\ref{sec:magnetic field} we explain the main features of our setup for generating a homogeneous magnetic field with a stability of a few milligauss. This is an important requirement to have precise control over intra- and inter-species scattering lengths tuned via magnetic FRs. 

\begin{figure}
	\centering
	\includegraphics[scale=0.6]{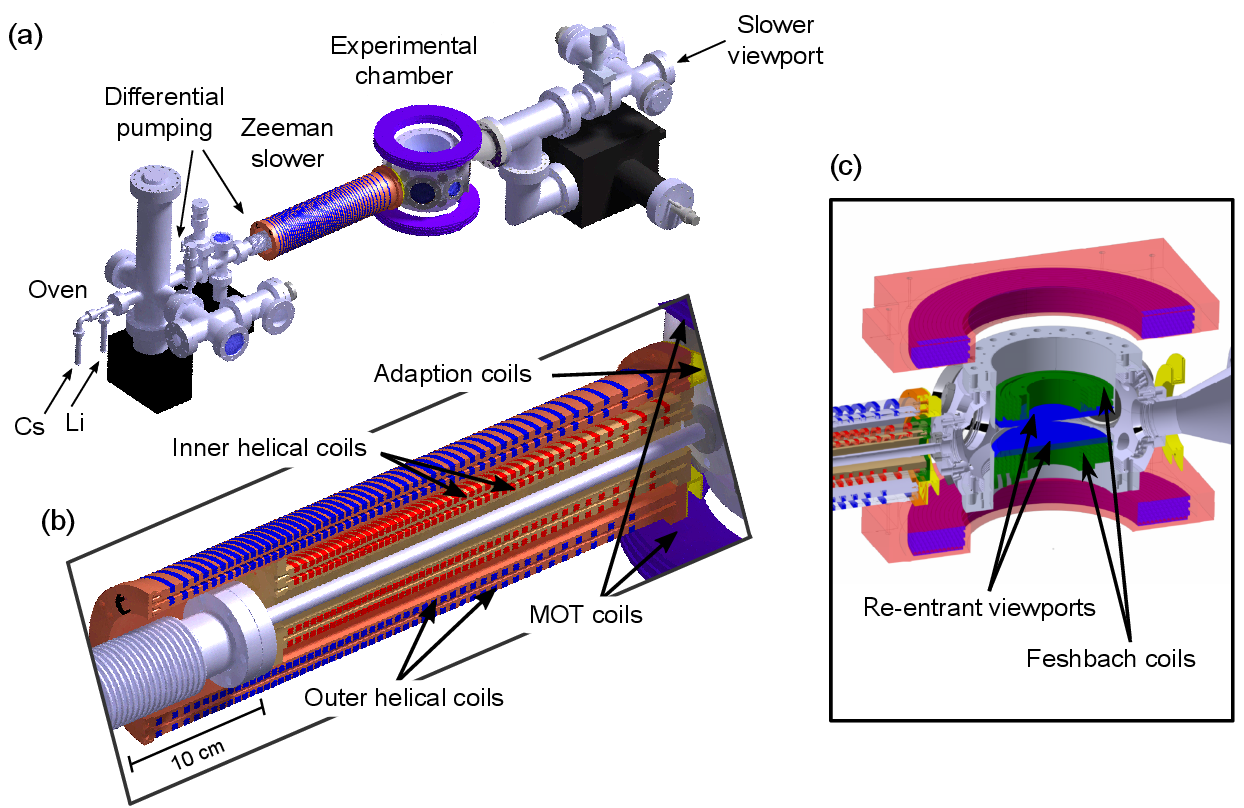}
	\caption{(a) Experimental apparatus: An atomic beam is generated in the oven chamber and decelerated by a double-species Zeeman slower for the loading of MOTs in the experimental chamber. Various ion pumps (black) in combination with differential pumping stages enable a pressure difference between the oven region and the experimental chamber. (b) Schematic cut through the Zeeman slower. Four helical coils, consisting of two inner ones (red) and two outer ones (blue), are shown. The fields for the last part of the Zeeman slower are generated by the radial fields of the MOT coils (violet) and a good matching of the fields is obtained by the adaption coil (yellow). (c) Schematic cut through the experimental chamber: Feshbach coils (green) are placed inside the recess of the re-entrant viewport. Pictures adapted from \citet{Repp2013PhD}.}
	\label{fig:Overview_Chamber}
\end{figure}

\section{Double-species Zeeman slower}\label{sec:slower}

A schematic overview of the core of our experimental apparatus for the production of ultracold $^6$Li-$^{133}$Cs mixtures is shown in Fig.~\ref{fig:Overview_Chamber}~(a). The two atomic beams are generated by a double-species effusive oven, decelerated by a double-species Zeeman slower consisting of four interleaving helical coils with variable pitch and captured in a magneto-optical trap inside the main experimental chamber by means of a sequential loading scheme. 
The oven region and the main chamber are connected by two differential pumping stages, allowing for pressure differences between the two parts. 
We refer the reader to \cite{Repp2013PhD} for a detailed description of our vacuum system.

In the following, in Sec.~\ref{subsec:oven} we give a brief description of our effusive oven, in Sec.~\ref{subsec:slower} we explain the design and the implementation of a double-species Zeeman slower and in Sec.~\ref{subsec:loading} we show the process of sequential loading of slow atoms of the two atomic species into a MOT. 

\subsection{Double-species effusive oven}\label{subsec:oven}

For the thermal source beam we use an effusive double-species oven with a similar design to the one presented by \citet{Stan2005}. It enables element selective control over the fluxes 
regardless of the difference between vapor pressures of $p_{\mathrm{Li}}= \SI{1e-5}{Torr} $ and $p_{\mathrm{Cs}}= \SI{4e-3}{Torr}$, at typical operation temperatures of $T_{\mathrm{oven, Li}}= \SI{625}{K}$ and $T_{\mathrm{oven, Cs}}= \SI{375}{K}$ \cite{Gehm2003,Steck2008}. 
The oven is divided into three parts maintained at different temperatures in order to have individual control on fluxes of the two atomic species, as explained in detail in \citet{Repp2013PhD}. The atomic beams leave the oven through a nozzle with a diameter of \SI{10}{mm} leading to atomic fluxes of $\Phi_{\mathrm{Li}}= \SI{5e15}{atoms/s}$ and $\Phi_{\mathrm{Cs}} = \SI{1e14}{atoms/s}$. Note that the velocity distribution in the atomic beams is strongly different for each species, due to the different temperature and mass for Li and Cs atoms.

\subsection{Design and implementation of the Zeeman slower}\label{subsec:slower}

The atomic beam produced by the effusive oven has mean velocities corresponding to \SI{1.48e3}{m/s} (Li) and \SI{244}{m/s} (Cs) that are a factor of 10 and 30 larger than the capture velocity of a MOT, respectively. One of the standard techniques to mitigate such a large gap is to use a Zeeman slower \cite{Phillips1982, MetcalfStraten1999}, where radiative light forces are used to decelerate the hot atoms while the changing Doppler shifts are compensated by the Zeeman shifts produced by a position-dependent magnetic field $B(z)$ along the atomic beam direction $z$. The required maximal magnetic field is expected to scale with $1/\sqrt{m}$ \cite{MetcalfStraten1999}, where $m$ is the atomic mass. Therefore, widely different magnetic field profiles are needed for decelerating either Cs or Li atoms. 

To create two different field profiles, we extended the approach presented by \citet{Bell2010} for a single species Zeeman slower where precise field profiles and fast switching times are generated by a single layer of windings with a variable pitch.
Our design is depicted in Fig.~\ref{fig:Overview_Chamber}~(b) and it consists of four interleaving layers of helical coils with radii of $R_\mathrm{1} = \SI{19.5}{mm}$, $R_\mathrm{2} = \SI{29.5}{mm}$, $R_\mathrm{3} = \SI{47.5}{mm}$ and $R_\mathrm{4} = \SI{57.5}{mm}$ and a total length of \SI{55}{cm}. Each layer is implemented as a single winding with a helical conductor profile of variable pitch. Moderate currents of \SI{30}{A} in the two outer layers generate the magnetic field profile for decelerating Cs atoms.
Much higher currents and differently shaped magnetic fields are reached  for cooling Li atoms by increasing the current in the two outer layers to \SI{75}{A} and simultaneously adding a second magnetic field created by the inner coils at the same current of \SI{75}{A}. 	
The small inductance due to a small number of windings allows for fast switching between loading Li or Cs. 
We measure switching times of the helical coils by recording the voltage response at the connectors while varying the control parameter of our power supplies. The 90\% level of the final voltage is reached within less than \SI{2}{ms} whereas, for switching off, the induced voltage vanishes faster than \SI{15}{ms}. Sub-millisecond timescales are also reachable for switching off the coils by electronically short-circuiting the coils. 

\begin{table}
	\caption{Design parameters used for simulating the optimal fields for the two species. The dimensionless deceleration $\eta$ is fixed to 0.5. The differences in $S_{\mathrm{0}}$ arise from different available laser powers on the experiment and the differences in $v^{\mathrm{in}}_{\mathrm{Cs,Li}}$ are assuming higher MOT capture velocities for Li than Cs due to the smaller mass.}\label{tab:design_par}%
	\begin{tabular}{llcc}
		\toprule
		\textbf{Design parameters}&&\textbf{Cs}&\textbf{Li} \\
		\midrule
		Current MOT Coils (A) &$I^{MOT}$&30&97.7 \\
		Current External Helical Coils ZS (A) &$I^{ext}$&30&75 \\
		Current Internal Helical Coils ZS (A) &$I^{int}$&0&75 \\
		Current Adaption Coils (A) &$I^{A}$&4.5&-1 \\
		Magnetic gradient MOT Coils (G/cm) & $\partial B/\partial z$  &9.5&31.0\\
		Saturation parameter & $S_{0}$  &10&2.5 \\
		Detuning slower laser (MHz) &$\delta/2\pi$  &-35&-70  \\
		Dimensionless acceleration & $\eta$ &0.5&0.5 \\
		Initial velocity (m/s) & $v^{in}$  &23&48 \\
		Minimal distance (cm) & $d^{min}$ &7.5&12.5\\
		\botrule
	\end{tabular}
\end{table}

The magnetic fields in the last part of the Zeeman slower are generated by the radial fields of the MOT coils \cite{Schunemann1998}. The radially symmetric quadrupole fields for the MOT are provided by two parallel coils consisting of 72 windings each. The magnetic fields required by design are generated by currents of $I_{\mathrm{Li}}= \SI{97.7}{A}$ and $I_{\mathrm{Cs}}= \SI{23.7}{A}$ for the Li and Cs configuration, respectively. The minimal radius of \SI{100}{mm} and the minimal distance from the chamber center of \SI{102}{mm} are given by the geometry of the experimental chamber.  

The perfect matching between the fields generated by the helical and the MOT coils is ensured by an additional thin adaption coil mounted in the overlap region between the Zeeman slower and the MOT coils at a distance of \SI{140}{mm} from the MOT center. This coil consists of two parallel stacks where each one is made out of 11 windings with a minimal radius of \SI{42}{mm}. The required magnetic fields are generated by a current of $I_\mathrm{Li}=\SI{5.3}{A}$ and $I_\mathrm{Cs} = \SI{-1.7}{A}$ for the Li and Cs configuration, respectively. 	
The smooth overlap between the magnetic fields is beneficial to minimize the transverse expansion of the atomic beam \cite{Joffe1993}, which is particularly critical for light elements like Li. Compared to other solutions \cite{Marti2010}, this design allows avoiding regions where the atomic beam is not decelerated and thus expands.

The required individual fields for decelerating Li and Cs atoms were calculated by studying the equations of motions via numerical simulation. 
The deceleration is provided by the dissipative light force $F_{\mathrm{D}}=\hbar k \gamma_{\mathrm{p}}$, where $k$ is the wave-vector of the light and $\gamma_{\mathrm{p}}$ is the excitation rate for a two level atomic system  \cite{MetcalfStraten1999}. At each position $F_{\mathrm{D}}$ has to be maintained constant and smaller than the limit-acceleration provided at saturation by the on-resonance photon scattering $F^\mathrm{max}_\mathrm{D}=\hbar k \gamma/2$, where $\gamma$ is the natural line-width of the transition.	
This requirement can be quantified with an dimensionless deceleration $\eta=F_\mathrm{D}/F^\mathrm{max}_\mathrm{D} = a/a_{\mathrm{max}}$ that must be kept $<1$ in order to fulfill the resonance condition at every position.

In Tab.~\ref{tab:design_par} the design parameters used for simulating the optimal fields for the two species are listed. For simplification, the deceleration process was inverted by virtually accelerating the two species separately towards the oven, starting at the MOT center with the initial velocities $v^{\mathrm{in}}_{\mathrm{Cs,Li}}$  equal to the capture velocities of $v^\mathrm{cap, MOT}_{\mathrm{Cs}}= \SI{23}{m/s}$ and $v^\mathrm{cap, MOT}_{\mathrm{Li}} = \SI{48}{m/s}$ of our Cs and Li MOT, respectively. The simulation started by first accelerating the atoms only by the radial fields of the MOT coils at the given currents. The design values for the  magnetic gradients of the MOT along the axial direction are $\partial {B}_{\mathrm{Cs}} /\partial {z}= \SI{9.5}{G/cm}$ and $ \partial {B}_{\mathrm{Li}} /\partial {z}= \SI{31.0}{G/cm}$, which enables the levitation of Cs atoms \cite{Weber2003a} while simultaneously loading a Li MOT. After a distance of about \SI{100}{mm}, where the chamber geometry allows for generating magnetic fields by additional coils, the photon scattering rate and thus the acceleration at each position afterwards was maximized in the simulation to $\eta \cdot a_{\mathrm{max}}$ via optimizing the Zeeman shift. 
For typical magnetic field values, the Zeeman shift of Li is larger than the hyperfine splitting and thus closed optical transitions within the Paschen-Back regime were used in the calculation. To maximize the atomic flux, the field was chosen to decelerate all three lowest $m_\mathrm{I}$ sub-levels simultaneously.

\begin{figure}
	\centering
	\includegraphics[scale=0.5]{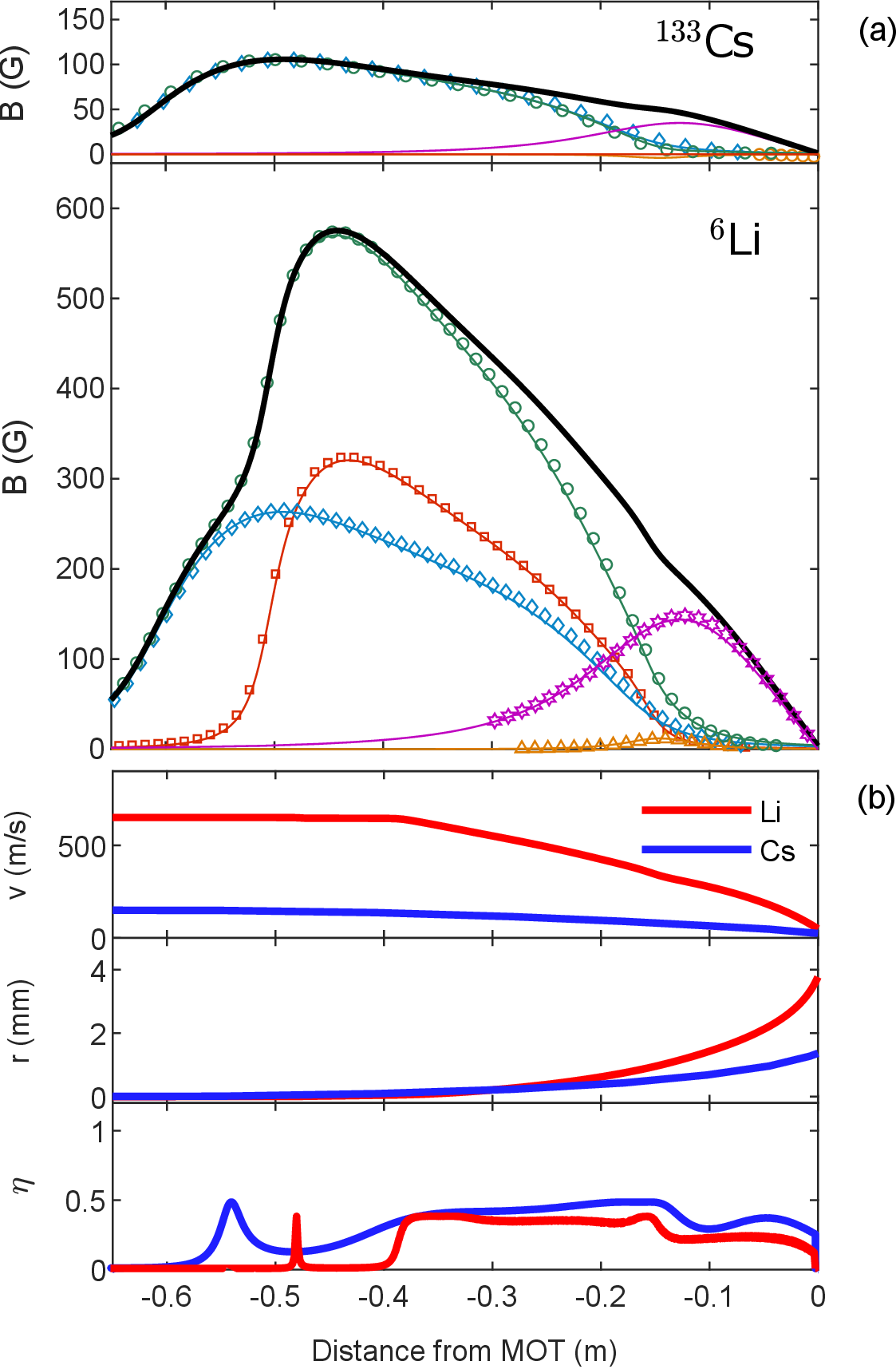}
	\caption{(a) Magnetic fields for the deceleration of Cs (upper panel) and of Li (lower panel). The measured field (green circles) is a sum of the fields from the inner helical coils (red squares), the outer helical coils (blue diamonds) and a small field from the adaption coil (orange triangles). The transverse expansion of the atomic beam is minimized by generating the last part of the total field for slowing via the radial field of the MOT coils (violet hexagrams). The colored lines represent the results from the calculation for the individual coils while the black lines depict the calculated total field of all coils. The only free parameters are the zero positions of the measurements. Data are taken from \citet{Repp2013PhD}. (b) Simulations of the deceleration process for Cs (blue) and Li (red). The longitudinal velocity (upper panel), the beam radius due to the transverse heating (middle panel) and the dimensionless acceleration $\eta$ (lower panel) are shown for atoms near the capture velocity of the Zeeman slower, corresponding to $v^\mathrm{cap, ZS}_{\mathrm{Li}} = \SI{650}{m/s}$ for Li ($m_{\mathrm{I}} = -1$) and to $v^\mathrm{cap, ZS}_{\mathrm{Cs}} = \SI{150}{m/s}$ for Cs ($m_{\mathrm{F}}=3$). Numerical data are obtained from a code adapted from \citet{Repp2013PhD}.}
	\label{fig:Measured_fields}
\end{figure} 

Fig.~\ref{fig:Measured_fields} (a) shows the calculated and obtained magnetic field profiles measured for both configurations with a standard Hall probe. A maximal discrepancy between the calculated and measured fields of only \SI{10}{G} and, within the tubes of the helical coils, relative magnetic field deviations of less than 5\% are measured. This has been possible thanks to the high accuracy of the CNC based manufacturing process for the helical coils. Indeed, the calculated profiles are directly milled into \SI{7}{mm} thick aluminium tubes by a CNC milling machine, where the shapes, obtained from the optimization and parametrized similar to \citet{Bell2010}, were directly used as an input. In the region outside of the tubes, relative discrepancies up to 20\% are measured. However, the independent control over the local field generated  by the adaption coils allows optimizing the magnetic field in this region of transition between Zeeman slower and MOT coils.

Fig.~\ref{fig:Measured_fields} (b) shows the longitudinal velocity, the beam radius due to the transverse heating and the dimensionless acceleration $\eta$ for atoms near the capture velocity of the Zeeman slower, corresponding to $v^\mathrm{cap, ZS}_{\mathrm{Li}} = \SI{650}{m/s}$ for Li ($m_{\mathrm{I}} = -1$) and to $v^\mathrm{cap, ZS}_{\mathrm{Cs}} = \SI{150}{m/s}$ for Cs ($m_{\mathrm{F}}=3$) calculated from the total field of all coils (black lines). One can notice that at the position of the junction between the helical and MOT coils, where the adaption coils are located, the deceleration remains constant around the design value of around 0.5 even inside the experimental chamber. This setting ensures an efficient deceleration until the atoms reach the MOT center.
As a consequence, the radial extension of the Li beam only reaches a radius of about \SI{4}{mm}, corresponding  to the extension of the MOT laser beams and guaranteeing an efficent loading into the MOT.

\subsection{Loading into the MOT}\label{subsec:loading}

\begin{figure}
	\centering
	\includegraphics[scale=0.35]{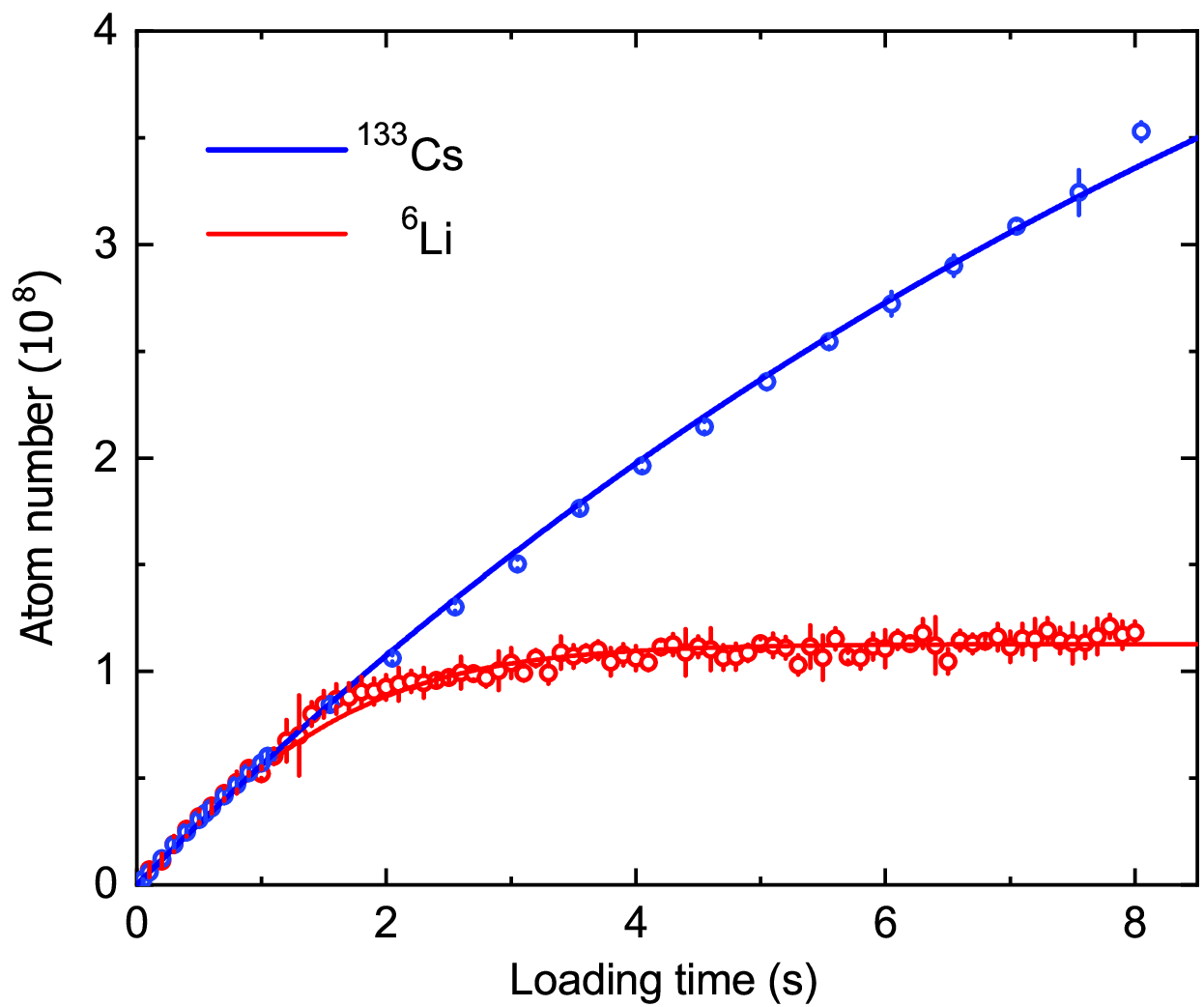}
	\caption{Time dependence of the number of Cs (blue) and Li (red) atoms loaded into their respective MOTs. Data (circles) are recorded by performing absorption imaging and the lines are the best fit to the models described in the text. A loading rate of about $\SI{6e7}{atoms/s}$ for both Li and Cs atoms is deduced by the fit. To resolve two-body effects, a higher number of data points is used in the Li measurement. Each point is the average of at least four measurements and the error bars represent the standard deviation. Data are taken from \citet{Lippi2024}.}
	\label{fig:Loadingrates_Slower}
\end{figure}

In order to test the performance of the Zeeman slower, loading curves for Li and Cs MOTs are recorded by using standard absorption imaging, as shown in Fig.~\ref{fig:Loadingrates_Slower} for $T_{\mathrm{oven}, \mathrm{Li}}= \SI{635}{K}$ and $T_{\mathrm{oven}, \mathrm{Cs}}= \SI{375}{K}$. In order to obtain fast cycle rates and an efficient sequential loading, a high loading rate is desired.

For Cs atoms, the measured loading curve is well reproduced by a exponential growth $N(t)=\frac{R_{\mathrm{Cs}}}{\alpha_{\mathrm{Cs}}}(1-e^{-\alpha_{\mathrm{Cs}} t})$ taking into account the  interplay between the increasing of the number of atoms via decelerated atoms at rate $R_{\mathrm{Cs}}$ and the density independent one-body loss mechanisms at rate $\alpha_{\mathrm{Cs}}$.  A typical loading curve for Cs results in a MOT loading rate of ${R}_{\mathrm{Cs}} = \SI{5.9(1)e7}{atoms/s}$ and with one-body loss rate of $\alpha_{\mathrm{Cs}}=\SI{0.09(1)}{s^{-1}}$. The 1/$e$-lifetime of approximately \SI{11}{s} is close to the vacuum-limited value, which is about a few tens of seconds \cite{Tran2022}.  For a typical MOT loading time of \SI{1}{s}, temperatures of ${T}_{\mathrm{Cs}}\approx T_{\mathrm{Doppler}} = \SI{125}{\micro K}$ are generally reached at this stage with corresponding atomic densities of $\SI{3.1e10}{cm^{-3}}$.

In the case of Li on the other hand, high densities result in non-negligible two-body losses induced by multiple scattering of photons and light-assisted collisions. Indeed, a maximal peak density of $\SI{7.2e10}{cm^{-3}}$ leads the atom number to saturate at the value of \num{1.2e8} for loading times longer than \SI{2}{s}.  Here, the evolution of the atom number is described by $\dot{N}(t)=R_{\mathrm{Li}}-\beta'_{\mathrm{Li}} N(t)^2$ with $\beta'_{\mathrm{Li}}=\beta_{\mathrm{Li}}/V$ where $\beta_{\mathrm{Li}}$ is the two-body loss coefficient and  
$V=(\sqrt{2 \pi} r)^3$ is the effective volume occupied by the cloud, where $r$ is the $1/e^2$ size of the cloud averaged in the three spatial dimensions.
The time evolution used for the fit is then given by $N(t)=\sqrt{R_{\mathrm{Li}}/\beta'_{\mathrm{Li}}}\tanh{(t \sqrt{R_{\mathrm{Li}}\beta'_{\mathrm{Li}}})}$ with a typical loading rate of ${R}_{\mathrm{Li}} = \SI{5.93(7)e7}{atoms/s}$. In agreement with what is reported in \cite{Shunemann1999, Schloder1999}, the two-body loss coefficient is $\beta_{\mathrm{Li}} = \SI{1.7(5)e-11}{cm^{3}}{s^{-1}}$. 
Typical achieved temperatures are around \SI{1.4}{mK}. 

\section{Dipole trap management}\label{sec:mixture}	

Simultaneous optical trapping of highly mass-imbalanced atomic species has to satisfy very different requirements for each of them, especially for reaching ultra-low temperatures. 
In Sec.~\ref{sec:Single-wavelength trapping} we present a single-wavelength trapping configuration for creating mixtures at temperatures as low as \SI{200}{nK}.
This scheme has been used for experiments presented in \cite{Repp_2013, Pires_2014, Ulmanis_2016, zhu2019, zhu2019a}. 
In Sec.~\ref{sec:Bichromatic trapping} we describe a bichromatic trapping scheme based on the method of species-selective trapping \cite{LeBlanc2007}, that allows realizing Li-Cs mixtures with temperature of \SI{100}{nK}. This scheme enabled the study of the role played by the boson-boson scattering length in the heteronuclear Efimov scenario described in \cite{Ulmanis_2016a, Haefner_2017}. 

\subsection{Single-wavelength trapping}\label{sec:Single-wavelength trapping}

An efficient transfer of laser-cooled atoms into an optical dipole trap requires a proper adaption of the trap depth to the temperature of the atomic cloud during the trap loading process. On the one hand, Cs is routinely laser-cooled to temperatures of ${T}_{\mathrm{Cs}} \approx \SI{1}{\micro K}$, but it suffers from severe three-body losses \cite{kraemer2006}. On the other hand,
the unresolved hyperfine structure of the excited $2^2\mathrm{P}_{3/2}$ state in Li prevents the application of simple sub-Doppler cooling schemes on the $\mathcal{D}_2$ transition and temperatures of standard Li MOTs are limited by the Doppler temperature ${T}_{\mathrm{Li}}\approx \SI{140}{\micro K}$, which require a much deeper dipole trap than Cs. The individual requirements during the transfer process can be satisfied by implementing  a sequential transfer scheme into spatially separated ODTs.

\begin{figure}
	\centering
	\includegraphics[scale=0.31]{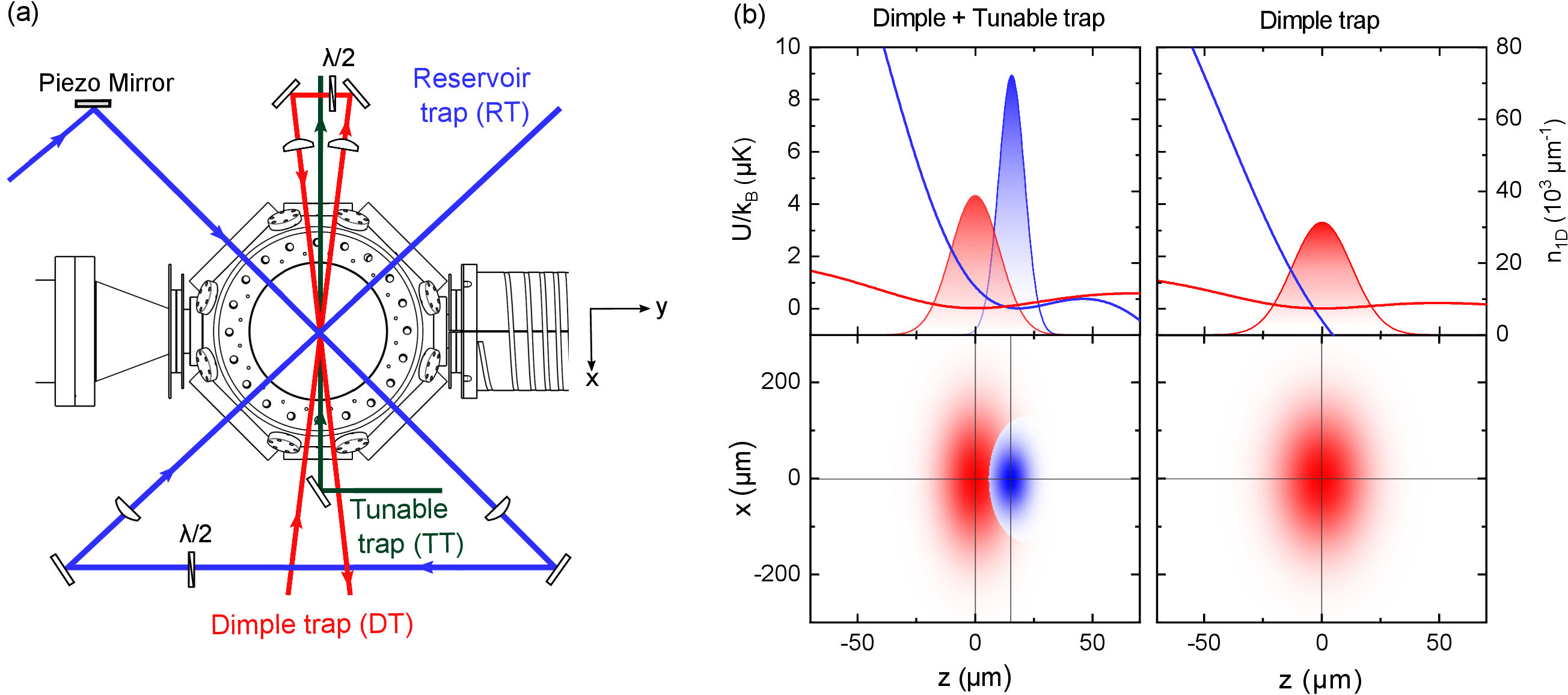}
	\caption{(a) Schematic overview of the optical dipole traps: reservoir trap (blue), dimple trap (red) and tunable trap (green). (b) Optical dipole potentials in presence of gravity in the z-direction for Cs (blue line) and Li (red line) at \SI{100} {nK} and their corresponding  atomic density distributions (blue and red faded areas: along the $z$-axis on the top panels, and in the $x$-$z$ plane on the low panels). On the left side, the potential provided by both DT and TT allows for a partial spatial overlap between the two clouds. Instead, the single-wavelength trapping provided by only the DT (right side) leads, for Cs, to a  completely tilted potential in which the atoms cannot be trapped. This picture is adapted from \citet{Ulmanis2015}.}
	\label{fig:potential}
\end{figure}

We use a combination of two different dipole traps as shown in Fig.~\ref{fig:potential}~(a), i.e. the reservoir trap (RT) and the dimple trap (DT). The RT at \SI{1064}{nm} is a crossed beam dipole trap under an angle of $90\degree$ with a beam waist of \SI{300}{\micro m}. This large optical dipole trap is provided by a \SI{55}{W} Nd:YAG solid-state laser and it is used for loading Cs atoms at trap depths on the order of \SI{10}{\micro K}. The DT is formed by two crossed linearly polarized beams at a crossing angle of $8.5\degree$  from a high-power fiber laser with a wavelength of \SI{1070}{nm}. At the maximal power of \SI{100}{W} per beam a maximal trap depth of \SI{1.2}{mK} for Li is achieved, which is necessary in order to load Li atoms from the MOT. The large trap frequencies on the order of \SI{10}{kHz} allow fast thermalization during loading and evaporative cooling. This trap is finally used for combined trapping of Li and Cs atoms.

In order to efficiently load the Li and Cs atoms to the ODT, additional cooling and state-manipulation schemes are applied. In short, each experimental cycle begins by loading a Cs MOT for a typical duration of \SI{1}{s} (see Sec.~\ref{subsec:loading}), followed by a \SI{35}{ms} long compression phase. Subsequently, a sub-Doppler cooling \cite{Drewsen1994} scheme is applied, resulting in a sample of \num{6e7} atoms where temperatures of around \SI{10 }{\micro K} are reached. At this point the RT is turned on at a power of \SI{25}{W}. While the dipole trap is on, Cs atoms are then prepared by means of the degenerate Raman-sideband cooling (DRSC)\cite{Vuletic1998,Kerman2000, Treutlein2001} in the lowest hyperfine state $\ket{F,m_{\mathrm{F}}}=\ket{3,3}$ of Cs, where two-body losses are suppressed. Waiting one quarter of the RT period, the atoms accumulate in the center of trap and by means of a subsequent \SI{1.5}{ms} DRSC pulse, the excess kinetic energy is removed. The magnetic fields as well as the beam powers of the Raman lattice and polarizer are ramped down in order to achieve temperatures of less than \SI{1}{\micro K} and a spin polarization of 90\%. By applying a homogeneous external magnetic field of $\approx \SI{4.5}{G}$, we keep Cs spin-polarized and obtain typical samples of  \num{4e7} atoms. 

At this stage, Li atoms are loaded in a spatially separated MOT for a duration of \SI{2.5}{s}. 
After a compression phase of the MOT,  \num{1.6e6} atoms are directly transferred into the DT, located $\approx\SI{1}{mm}$ away from the position of the Cs atoms confined in the RT. Li atoms are optically pumped into the $F=1/2$ hyperfine manifold, where they equally populate the two sub-levels with $m_{\mathrm{F}}=\pm1/2$.

Forced evaporative cooling is performed for 3 s on the two separated atomic samples at a magnetic  field offset of $\approx \SI{896}{G}$ where the large intraspecies scattering lengths $a_{\mathrm{Li}}\approx \SI{-8000}{a_0}$ and $a_{\mathrm{Cs}}\approx \SI{400}{a_0}$ ensure thermalization of both species. The two traps are then overlapped within \SI{1}{s} by moving the beams of the RT of $\sim1$ mm, by a piezo-driven mirror, at a magnetic field of \SI{907}{G}, where the large scattering length between the Li$\ket{1/2,1/2}$ and Cs$\ket{3,3}$ states leads to sympathetic cooling of the Cs atoms. 

At this point, to prepare Li-Cs mixtures at final temperatures down to \SI{200}{nK}, the power of the RT is linearly ramped down and turned off within \SI{1.5}{s}. The mixture is therefore confined in the single-wavelength DT. A Li spin-state is selected by removing the other one with a \SI{80}{\micro s} long resonant light pulse. The final number of Li and Cs atoms, temperatures, and trapping frequencies are subject to the exact trapping conditions and can be tuned within a certain range. Typical atom numbers at \SI{430}{nK} are \num{5e4} for Cs and \num{3e4} for Li, with trapping frequencies of $\omega_{Cs}=2\pi\times(12,173,162)$ Hz and $\omega_{Li}=2\pi\times(44,388,388)$ Hz.

\subsection{Bichromatic trapping}\label{sec:Bichromatic trapping}	

The challenge to further decrease the temperature of the mixture in a single-wavelength ODT arises from the significantly different gravitational potentials due to the mass imbalance between Li and Cs. The monochromatic trapping is indeed possible only as long as the dipole potential dominates over the gravitational potential, preventing to perform further forced evaporation.

At the wavelength of the DT, the difference in the atomic polarizability of a factor of ${\alpha_{\mathrm{Cs}}}/{\alpha_{\mathrm{Li}}}\approx 4$, results in the same factor in the potential depths such that $U^{\mathrm{Cs}}_{\mathrm{dip}} \approx 4U^{\mathrm{Li}}_{\mathrm{dip}}$. Considering a harmonic trap confinement, the gravitational sag is given by $z_\mathrm{0} \propto g/\omega_\mathrm{z}^2\propto m/\alpha$, where $g$ is the gravitational acceleration, $\omega_z$ is the trapping frequency along the vertical direction, $m$ is the atomic mass and $\alpha$ is the atomic polarizability. For a mass ratio of 22, $z^{\mathrm{Cs}}_{0}/z^{\mathrm{Li}}_{0}=\frac{m_{\mathrm{Cs}}}{m_{\mathrm{Li}}} \frac{\alpha_{\mathrm{Li}}}{\alpha_{\mathrm{Cs}}} \approx 5$. 
Therefore, to mitigate the large differential gravitational sag, a further trapping potential at a wavelength with a larger Cs-Li polarizability ratio is implemented. 
This ODT, called tunable trap (TT), is based on the method of species-selective trapping \cite{LeBlanc2007} and it provides individual control over the potentials of each species. To implement different possible trapping scenarios, the trap is designed to be wavelength tunable from \SI{700}{nm} to \SI{1030}{nm}. The desired light is supplied by a titanium-sapphire laser (Coherent MBR-100 pumped by a Coherent-Verdi V-18), delivering up to \SI{3}{W} optical power. The TT is implemented by a single focused beam along the axis of the cigar-shaped DT (see Fig. \ref{fig:potential} (a)) with a waist of \SI{60}{\micro m} and in combination with the DT allows nearly harmonic potentials.
We choose a wavelength of \SI{921.1}{nm} for the TT, where the polarizability for Cs (\SI{4060}{a.u.}) is of a factor 12 times greater than the one for Li (\SI{345}{a.u.}). This allows increasing the trapping potential restoring the confinement against gravity for Cs atoms, while the potential depth of Li only experiences minor changes. This wavelength was optimized experimentally as a compromise between differential polarizabilities and sufficiently small heating rates for Cs. The relative trap depth and the residual gravitational sag can be controlled by relative shifts and intensities of the two dipole traps.

For the preparation of Li-Cs mixtures at temperatures of \SI{100}{nK}, once RT and DT are superimposed and before switching off the RT and performing further forced evaporation, the TT is turned on to \SI{32}{mW} within \SI{400}{ms}. Forced evaporation is performed by ramping down the RT within \SI{1}{s}, leaving the atoms trapped only by DT and TT. The last evaporation phase is performed at \SI{920}{G} where the DT power is linearly decreased to \SI{42}{mW} while simultaneously increasing the power of the TT to \SI{37}{mW}. This procedure results in a good spatial overlap and similar trap depths for Li and Cs during evaporation, as shown in Fig.~\ref{fig:potential}~(b). 
In this case, we obtain a confining potential for both species in the direction of gravity. On the other hand, a pure single wavelength trap leads to a drastic reduction in the effective trap depth for Cs and then to an effective shift in the position of the minimum.
We can further notice that the position of Li atoms is almost unchanged and thus the overlap of the two species is reduced. 
With this scheme, non-degenerate samples of \num{1e4} Cs atoms and \num{7e3} Li atoms, with trapping frequencies of $\omega_{\mathrm{Cs}}=2\pi\times(5.7,115,85)$ Hz and $\omega_{\mathrm{Li}}=2\pi\times(25,160,180)$ Hz, are obtained. Indeed, for a final temperature of \SI{120}{nK} a local $T/T_{\mathrm{F}}\geq2$ for Li and $T/T_{\mathrm{C}}\geq3$ for Cs is calculated, where $T_{\mathrm{F}}$ and $T_{\mathrm{C}}$ are the Fermi temperature and the critical temperature for Bose-Einstein condensation in three dimensions, respectively.

For our experiments on the Efimov effect~\cite{Ulmanis_2016a}, a trap depth of approximately \SIrange[range-phrase=--]{500}{700}{nK} and a finite spatial overlap of the atomic clouds are required.
This situation can be achieved by a vertical displacement of the TT of \SI{25}{\micro m} and powers of the DT and TT trap of about \SI{40}{mW}. The differential sag amounts to \SI[separate-uncertainty = true]{16(5)}{\micro m} yielding an estimated spatial overlap of 45$\%$  to the case where the central position of the distributions is the same. This scenario is shown in Fig.\ref{fig:potential}~(b).  The restoring of a partial overlap of the two clouds in the \SI{100}{nK} regime enables us to reveal the second excited Efimov resonance as shown in Fig.~\ref{fig:3bodyLoss}~(b) \cite{Ulmanis_2016a} with an excellent matching of the experimental data with the theory model \cite{Petrov_2015}. 
The obtained enhancement of the three-body loss rate $L_3$ in proximity of the second excited Efimov state at temperatures of \SI{120}{nK} confirmed the previous signature measured for \SI{450}{nK} in a single-wavelength configuration where this feature was only weakly visible \cite{Pires_2014}. 

\section{Magnetic field control}\label{sec:magnetic field}	

Magnetic field stability has fundamental relevance for all the experiments where it is important to control and tune the interactions between the atoms close to Feshbach resonances. 
Homogeneity and long-term stability of our magnetic fields allow us to exquisitely tune interactions for studying the Efimov scenario \cite{Pires_2014,Ulmanis_2016, Ulmanis_2016a, Haefner_2017} and to perform high-resolution atom-loss spectroscopy \cite{zhu2019, zhu2019a, Gerken2019}.

In the following, we explain the main characteristics of our setup for generating homogeneous stable magnetic fields highlighting the most important properties: homogeneity in Sec.~\ref{sec:Homogeneity}, calibration and long-term stability in Sec.~\ref{sec:Calibration}. The obtained magnetic field resolution is discussed in Sec.~\ref{sec:Summary}.

\begin{figure}
	\centering
	\includegraphics[scale=0.35]{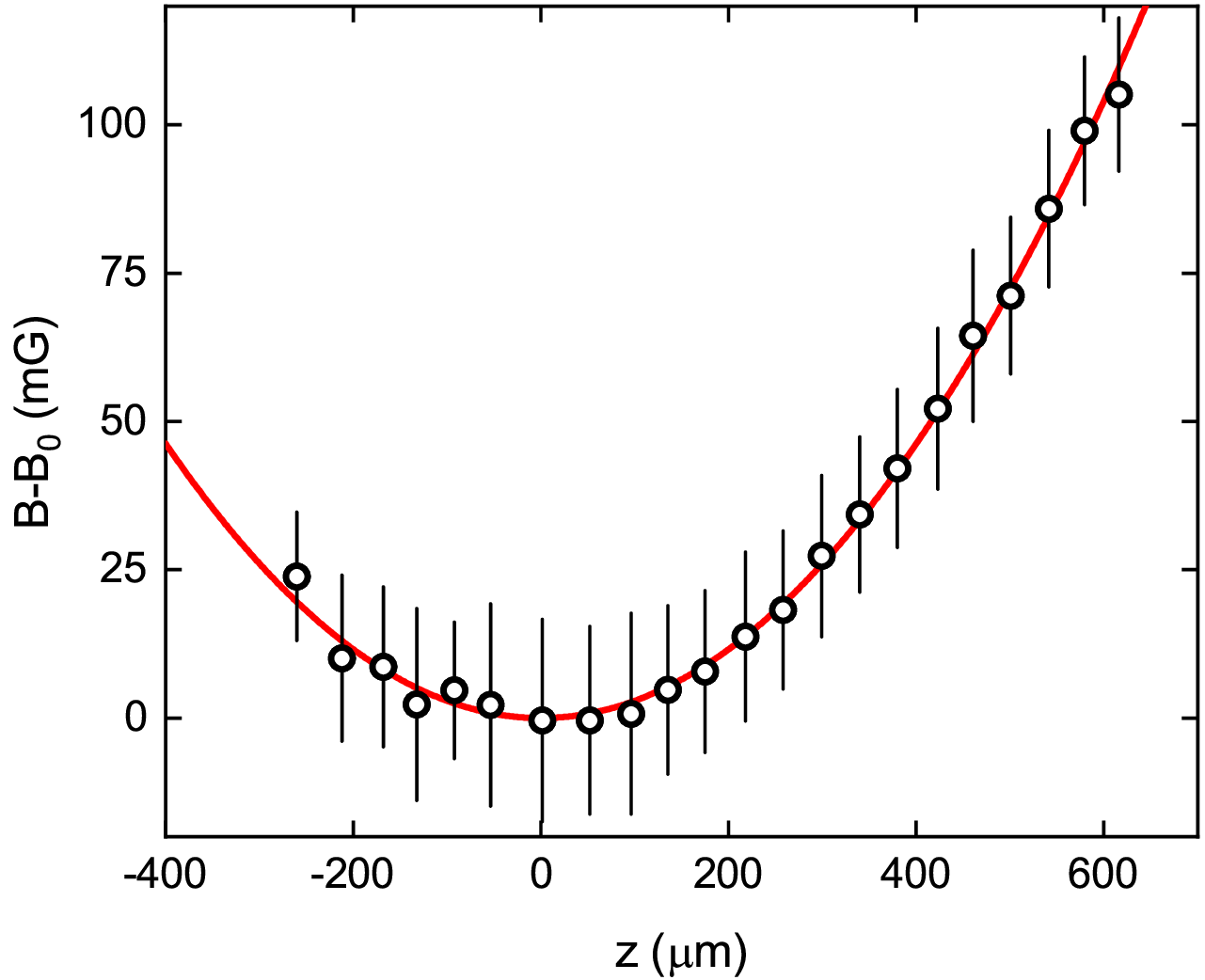}
	\caption{Tomography of the magnetic field produced by the Feshbach coils along the \textit{z}-axis (direction of gravity). The magnetic field is measured by microwave spin flip on Cs atoms confined in the RT for different positions (open circles) and is given as deviation from its maximum value $B_0 = \SI{1074.95}{G}$. The red line is a parabolic fit and yields a curvature of 0.0270(4) cm$^{-2}$. Each point is the result of the average of more than thirty measurements taken with a binning of 5 pixels ($\approx$ 40 $\mu$m).The error bars correspond to fluctuations over a period of 7 hours. Data adapted from \citet{Gerken2022}.}
	\label{fig:tomography}
\end{figure}

\subsection{Homogeneity}\label{sec:Homogeneity}

A homogeneous magnetic field is generated by a couple of coils mounted in Helmholtz configuration within the re-entrant viewports of the main chamber as shown in Fig. \ref{fig:Overview_Chamber}~(c). The design of the coils consists of 4$\times$6 windings, with a minimal radius of \SI{39.1}{mm} and a distance from the center of the chamber of \SI{19.5}{mm}. They generate magnetic fields up to \SI{1350}{G} corresponding to the maximum available current of \SI{400}{A}. The temperature of the coils is stabilized by cooling water supplied from a custom made industrial cooling unit to $\Delta T<\SI{0.3}{K}$ and the current, measured by a current transducer, is actively stabilized with a PID controller.
Nevertheless, the placement of the coils in the re-entrant viewports prevents the realization of a perfect Helmholtz configuration leading to a finite curvature of the magnetic field at the center of the chamber. In order to detect the magnetic field inhomogeneity along the $z$ axis we adopted a tomographic approach.
Cs atoms are prepared in the RT in $\ket{F,m_{\mathrm{F}}}=\ket{3,3}$ where, depending on its detuning, a \SI{35}{\micro s} microwave pulse selectively excites them at different positions to the $\ket{4, 4}$ state.
The center position of the atom cloud can be directly recorded by performing a gaussian fit for different microwave detuning and then calculating the corresponding magnetic field using the Breit-Rabi formula. To have a complete tomography, we move the cloud by means of the piezo-mirror over a region of about \SI{1}{mm} (see Sec.~\ref{sec:Single-wavelength trapping}).
The variation of the measured magnetic field from the bias field $B_0 = \SI{1074.95}{G}$ is shown in Fig.~\ref{fig:tomography} and can be reproduced by a parabolic function $B(z)-B_0=B_0 c_z z^2$, where $c_z = 0.0270(4)$ cm$^{-2}$ is the obtained coefficient for the curvature along $z$-direction. Each fit result is averaged over 5 pixels ($\approx$ 40 $\mu$m), and fluctuations across multiple measurements cause a shift to higher magnetic fields. This shift, however, is irrelevant for estimating the finite curvature of the coils and results in an uncertainty of \SI{16}{mG} over several hours. 
The obtained profile shows that a displacement of a few hundred micrometers from the center results in a magnetic field variation of only a few milligauss. For a typical experiment with a Li-Cs mixture  at \SI{400}{nK}, thermal cloud sizes are $\sigma_{Li}=\qtylist[list-units = bracket, list-final-separator = {, }]{140;10;10}{\micro m}$ and $\sigma_{Cs}=\qtylist[list-units = bracket, list-final-separator = {, }]{70;5;6}{\micro m}$, respectively, and the resulting variation in the $z$-direction is less than \SI{1}{mG}. 
The upper limit corresponds to the variation obtained for Li in the direction of the major elongation of the cloud. This is estimated taking into account that $B(x)-B_0=B_0 c_x \sigma_{Li,x}^2$ with $c_x = c_z/2 = \SI{0.0134(2)}{cm^{-2}}$ and a large bias field of \SI{1000}{G}.
For a typical size of $\sigma_{Li,x} \approx \SI{150}{\micro m}$ the magnetic field variation is about \SI{3}{mG}. 

\subsection{Calibration and stability}\label{sec:Calibration}
Calibration of the magnetic field is performed using radio-frequency spectroscopy on Li. For these measurements, a spin-polarized Li sample is prepared in the hyperfine state $\ket{1/2,1/2}$ and transferred to $\ket{1/2,-1/2}$ by applying a rectangular radio-frequency pulse. The remaining number of Li atoms in the state $\ket{1/2,1/2}$ is recorded and fitted with a sinc-shape function. The magnetic field strength is obtained using the Breit-Rabi energy difference. The typical statistical error in the determination of the resonance frequency is on the \SI{1}{Hz} level corresponding to a magnetic field accuracy of less than \SI{1}{mG} \cite{Hafner2017a}. By repeating the measurements for different values of the control parameter, a calibration of the magnetic field is obtained. This procedure leads to a maximal uncertainty of \SI{9}{mG} \cite{Ulmanis2015}.

At this  point, long-term stability has to be taken into consideration. 
During a time of several days, the temperature of the magnetic field coils and the position of the optically trapped atoms change, leading to a variation of the magnetic field strength.  The lowest uncertainty is achieved when the experiment reaches thermal equilibrium, which, if initialized from a completely cold state, takes about one day. In such a condition, long-term stability measurements of the magnetic field over time, as reported in \cite{Ulmanis2015}, show a maximum standard deviation from the mean of \SI{8}{mG} over a period of 24 hours. 

Another source of fluctuation is given by the length of the experimental sequence. The accumulated ohmic heating can deform the geometry of the coils and thus can lead to a different absolute applied magnetic field as a function of the experimental sequence length. This effect yields an uncertainty of \SI{8}{mG}, quantified from the comparison between the magnetic field calibration immediately after the sample preparation and after holding for additional \SI{3}{s} \cite{Ulmanis2015}.

\subsection{Resolution}\label{sec:Summary}

\begin{figure}
	\centering
	\includegraphics[scale=0.35]{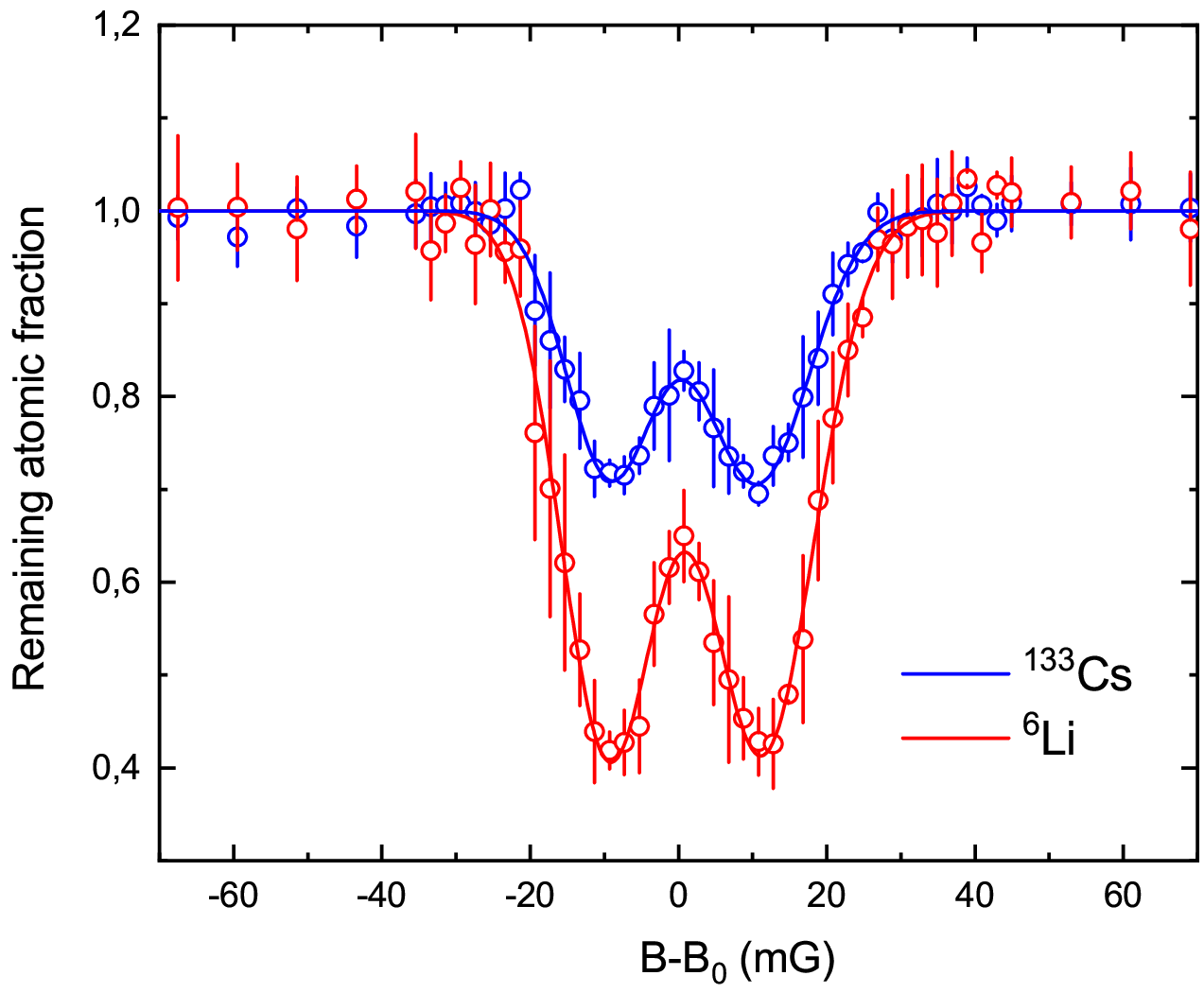}
	\caption{ Feshbach spectroscopy of the Li-Cs \textit{p}-wave FRs at \SI{708}{G}. The remaining fraction of Li atoms in the $\ket{1/2,-1/2}$ state (red) and of Cs atoms in the $\ket{3,3}$ state (blue) after an optimized holding time of \SI{1}{s} with respect to the magnetic field are shown. $B_0$ has been chosen arbitrarily to be at half distance between the Cs peaks and corresponds to \SI{708.890}{G}. Each point is the average of at least four measurements and the error bars represent the standard deviation. The curves are fits of a sum of two Gaussian functions. Data are taken from \citet{Hafner2017a}.}
	\label{fig:pwaves}
\end{figure}

Residual magnetic field gradient, calibration, long-term stability and length of the experimental sequence 
are the main sources of systematic error in the magnetic field determination. We assume that all these uncertainties are uncorrelated and by adding them up quadratically we obtain a total systematic error that amounts to \SI{16}{mG}. Such a good magnetic field stability permits us to have high resolution in atom-loss spectroscopy, as explained in the following.

The scattering length behavior close to a Feshbach resonance is given by $a(B)=a_{bg}(1-\frac{\Delta B}{B-B_0})$, where $a_{bg}$ is the background scattering length, $B_0$ is the pole of the resonance, and $\Delta B$ is its width. The maximum scattering length  $a_{max}$ that can be resolved is $a_{max}\approx a_{bg}\frac{\Delta B}{\delta B}$, where $\delta B$ is the total magnetic field uncertainty. 
For the Li-Cs case, using $\Delta B=$ 60 G, $a_{bg}=$ 30 a$_0$ \cite{Repp_2013} and an uncertainty of $\delta B=\SI{16}{mG}$, results in $a_{max}\approx10^5 a_0$. Therefore, it is evident that a stability on the order of a few tens of milligauss is necessary to resolve scattering lengths close to the pole of the resonance. 

A proof of the high magnetic field resolution achievable with this setup is obtained by performing atom-loss spectroscopy on magnetically induced \textit{p}-wave FRs in the Li-Cs mixture, as in the experiments reported in \cite{zhu2019, zhu2019a} and also shown in Fig. \ref{fig:pwaves}. The measured sample consists of a mixture at \SI{430}{nK} as described in Sec.~\ref{sec:Single-wavelength trapping}. The samples are held at different magnetic fields for an optimized holding time of \SI{1}{s} and FRs are identified as simultaneously enhanced atom losses in Li$\ket{1/2,-1/2}$ and Cs$\ket{3,3}$. A doublet close to \SI{708.890}{G} is observed. The magnetic field is randomly sampled and the step size is reduced from \SI{8}{mG} to \SI{2}{mG} around the loss features. Here, a splitting in the loss feature of \SI{20}{mG} was measured. 
Splittings of even \SI{6}{mG} have been resolved in the same kind of measurement in a spin-polarized pure $^6$Li sample close to a \textit{p}-wave resonance at \SI{185}{G} \cite{Gerken2019}.

\section{Conclusion}\label{sec:conclusion}

We report on the experimental apparatus and production of ultracold thermal Fermi-Bose mixtures of $^6$Li and $^{133}$Cs atoms, which have been employed in high-resolution atom loss spectroscopy experiments to investigate the heteronuclear Efimov effect.
A design consisting of four interleaving layers of coils is set up, which provides optimized fields for either cooling Li or Cs, allowing for efficient sequential loading of both species into their respective MOTs, with switching times of a few milliseconds between the two field configurations.  
The performance of our system is demonstrated by investigating MOT loading dynamics, resulting in a loading rate of about \SI{6e7}{atoms/s} for both species. This approach can be easily adapted to the efficient Zeeman slowing of other species with largely different masses. 

In addition, we demonstrate that the bichromatic trapping scheme presented allows for the preparation of an optically trapped Fermi-Bose mixture of Li and Cs atoms at a temperature of $T= \SI{100}{nK}$. This enabled us to resolve the challenging-to-detect enhancement of the three-body rate coefficient $L_3$ in proximity of the second excited Efimov state \cite{Ulmanis_2016a}. 
This was also achieved thanks to the highly stable magnetic fields. We demonstrate stability of a few tens of milligauss and estimate that this enables us to resolve Li-Cs scattering lengths up to $a_{max}\approx10^5 a_0$. 
These properties allow us to perform high-resolution atom-loss spectroscopy and to detect a splitting of up to \SI{20}{mG} in a $^6$Li-$^{133}$Cs mixture \cite{zhu2019, zhu2019a} and \SI{6}{mG} in a spin-polarized sample of $^6$Li \cite{Gerken2019}.

However, the trapping and evaporative cooling scheme presented here is not designed to produce double degenerate mixtures \cite{Silber2005, Hansen2011, DeSalvo2017, Chen2023}, which is of interest, for example, for studying signatures of few-body physics in many-body energy spectra, as in the polaron scenario \cite{Levinsen2015, Sun2017, Sun2017a}. Since in the degenerate regime the sizes of the clouds are on the order of the differential gravitational sag reported here, an improved experiment requires a scheme to combine the two species that allows for a better cancellation of this sag.

In the new generation of this experiment two alternative strategies have been implemented to tackle this challenge \cite{Lippi2024}. The first is to use a bichromatic trapping approach, similar to that presented here, but with the wavelength of the Ti:Sa laser set at the Cs tune-out wavelength, i.e. 880.25 nm \cite{Arora2011, Ratkata2021}. This scheme, combined with the implementation of a device to mechanically adjust the focus position of the trap at 880.25 nm, has allowed trapping of a degenerate Li Fermi Sea within a Cs thermal cloud without introducing additional confinement to Cs and cancelling out the differential gravitational sag. However, in this scheme the Cs atoms undergo spontaneous Raman scattering, which limits the timescales of the experiment that can be performed in such a configuration \cite{Lippi2024}. The second strategy, similar to the one implemented in \cite{DeSalvo2017, Chen2023}, is to use single wavelength trapping combined with a magnetic field gradient. 
In particular, if the force resulting from the gradient is oriented in the same direction of gravity, it pushes both Li and Cs in the same direction, resulting in a tilt of the trapping potential that is more pronounced for Li than for Cs. As in this configuration the shape and gravitational sag of the Li cloud  closely resembles the potential felt by Cs atoms, the differential gravitational sag can be effectively canceled \cite{Lippi2024}. 
The realization of heteronuclear systems of impurities coupled to a quantum bath, as well as double degenerate mixtures \cite{DeSalvo2017, Chen2023}, opens up to the study of a wide variety of many-body phenomena, ranging from Bose- and Fermi-polaron physics in heteronuclear mixtures \cite{Hu2016, Kohstall2012, Cetina2015, Cetina2016, Yan2020a, Fritsche2021a, Baroni2023} to the study of phase diagrams \cite{Duda2023}, collective excitations \cite{DeSalvo2019, Patel2023, Yan2024a} and droplet formation \cite{Rakshit2019a, Rakshit2019} in double degenerate Bose-Fermi mixtures.

\bmhead{Acknowledgements}
We acknowledge S. Schmidt, R. M\"uller, R. Heck, A. Arias and K. Meyer for their contribution during the different phases of building up and optimization of the experimental apparatus. Moreover, we acknowledge the institute's mechanical and electrical workshops and construction office for enabling the realization of the design of the whole machine. We are grateful to the members of the groups of S. Jochim, M. Oberthaler and R. Grimm for fruitful discussions. This work is supported by DFG/FWF Research Unit FOR2247 under Project No. WE2661/11-1, by the Deutsche Forschungsgemeinschaft (DFG, German Research Foundation) Project No.  273811115 - SFB 1225 ISOQUANT, by the Deutsche Forschungsgemeinschaft (DFG, German Research Foundation) under Germany´s Excellence Strategy EXC 2181/1 - 390900948 (the Heidelberg STRUCTURES Excellence Cluster) and by the Heidelberg Center for Quantum Dynamics. E.K. acknowledges the financial support by the Baden-W\"urttemberg Stiftung through the Eliteprogramme for Postdocs. E.L., M.R. and T.K. acknowledge support by the IMPRS-QD. 

\section*{Declarations}

\bmhead{Competing interests} The authors declare no competing interests.

\bmhead{Author contribution} E.L., S.H., M.Re. wrote the manuscript with the contribution of all authors; E.L., M.G., M.Re., S.H., J.U., B.Z., E.K., B.T. performed the measurements and all authors participated in the analysis of the data; L.C. and M.W. supervised the project.

\bibliography{manuscript_v2}


\begin{thebibliography}{82}
\ifx \bisbn   \undefined \def \bisbn  #1{ISBN #1}\fi
\ifx \binits  \undefined \def \binits#1{#1}\fi
\ifx \bauthor  \undefined \def \bauthor#1{#1}\fi
\ifx \batitle  \undefined \def \batitle#1{#1}\fi
\ifx \bjtitle  \undefined \def \bjtitle#1{#1}\fi
\ifx \bvolume  \undefined \def \bvolume#1{\textbf{#1}}\fi
\ifx \byear  \undefined \def \byear#1{#1}\fi
\ifx \bissue  \undefined \def \bissue#1{#1}\fi
\ifx \bfpage  \undefined \def \bfpage#1{#1}\fi
\ifx \blpage  \undefined \def \blpage #1{#1}\fi
\ifx \burl  \undefined \def \burl#1{\textsf{#1}}\fi
\ifx \doiurl  \undefined \def \doiurl#1{\url{https://doi.org/#1}}\fi
\ifx \betal  \undefined \def \betal{\textit{et al.}}\fi
\ifx \binstitute  \undefined \def \binstitute#1{#1}\fi
\ifx \binstitutionaled  \undefined \def \binstitutionaled#1{#1}\fi
\ifx \bctitle  \undefined \def \bctitle#1{#1}\fi
\ifx \beditor  \undefined \def \beditor#1{#1}\fi
\ifx \bpublisher  \undefined \def \bpublisher#1{#1}\fi
\ifx \bbtitle  \undefined \def \bbtitle#1{#1}\fi
\ifx \bedition  \undefined \def \bedition#1{#1}\fi
\ifx \bseriesno  \undefined \def \bseriesno#1{#1}\fi
\ifx \blocation  \undefined \def \blocation#1{#1}\fi
\ifx \bsertitle  \undefined \def \bsertitle#1{#1}\fi
\ifx \bsnm \undefined \def \bsnm#1{#1}\fi
\ifx \bsuffix \undefined \def \bsuffix#1{#1}\fi
\ifx \bparticle \undefined \def \bparticle#1{#1}\fi
\ifx \barticle \undefined \def \barticle#1{#1}\fi
\bibcommenthead
\ifx \bconfdate \undefined \def \bconfdate #1{#1}\fi
\ifx \botherref \undefined \def \botherref #1{#1}\fi
\ifx \url \undefined \def \url#1{\textsf{#1}}\fi
\ifx \bchapter \undefined \def \bchapter#1{#1}\fi
\ifx \bbook \undefined \def \bbook#1{#1}\fi
\ifx \bcomment \undefined \def \bcomment#1{#1}\fi
\ifx \oauthor \undefined \def \oauthor#1{#1}\fi
\ifx \citeauthoryear \undefined \def \citeauthoryear#1{#1}\fi
\ifx \endbibitem  \undefined \def \endbibitem {}\fi
\ifx \bconflocation  \undefined \def \bconflocation#1{#1}\fi
\ifx \arxivurl  \undefined \def \arxivurl#1{\textsf{#1}}\fi
\csname PreBibitemsHook\endcsname

\bibitem[\protect\citeauthoryear{Greene}{2010}]{Greene2010}
\begin{barticle}
\bauthor{\bsnm{Greene}, \binits{C.H.}}:
\batitle{{Universal insights from few-body land}}.
\bjtitle{Physics Today}
\bvolume{63}(\bissue{3}),
\bfpage{40}--\blpage{45}
(\byear{2010})
\doiurl{10.1063/1.3366239}
\end{barticle}
\endbibitem

\bibitem[\protect\citeauthoryear{Blume}{2012}]{Blume2012}
\begin{barticle}
\bauthor{\bsnm{Blume}, \binits{D.}}:
\batitle{{Few-body physics with ultracold atomic and molecular systems in
  traps}}.
\bjtitle{Reports on Progress in Physics}
\bvolume{75}(\bissue{4}),
\bfpage{046401}
(\byear{2012})
\doiurl{10.1088/0034-4885/75/4/046401}
{\href{https://arxiv.org/abs/1111.0941}{{arXiv:1111.0941}}}
\end{barticle}
\endbibitem

\bibitem[\protect\citeauthoryear{Frederico et~al.}{2012}]{Frederico2012}
\begin{barticle}
\bauthor{\bsnm{Frederico}, \binits{T.}},
\bauthor{\bsnm{Delfino}, \binits{A.}},
\bauthor{\bsnm{Tomio}, \binits{L.}},
\bauthor{\bsnm{Yamashita}, \binits{M.T.}}:
\batitle{{Universal aspects of light halo nuclei}}.
\bjtitle{Progress in Particle and Nuclear Physics}
\bvolume{67}(\bissue{4}),
\bfpage{939}--\blpage{994}
(\byear{2012})
\doiurl{10.1016/j.ppnp.2012.06.001}
\end{barticle}
\endbibitem

\bibitem[\protect\citeauthoryear{Wang et~al.}{2013}]{Wang2013}
\begin{bchapter}
\bauthor{\bsnm{Wang}, \binits{Y.}},
\bauthor{\bsnm{D’Incao}, \binits{J.P.}},
\bauthor{\bsnm{Esry}, \binits{B.D.}}:
\bctitle{Ultracold few-body systems}.
In: \beditor{\bsnm{Arimondo}, \binits{E.}},
\beditor{\bsnm{Berman}, \binits{P.R.}},
\beditor{\bsnm{Lin}, \binits{C.C.}} (eds.)
\bbtitle{Advances in Atomic, Molecular, and Optical Physics}
vol. \bseriesno{62},
pp. \bfpage{1}--\blpage{115}.
\bpublisher{Academic Press}, \blocation{???}
(\byear{2013}).
\doiurl{10.1016/B978-0-12-408090-4.00001-3}
\end{bchapter}
\endbibitem

\bibitem[\protect\citeauthoryear{Efimov}{1970}]{Efimov1970}
\begin{barticle}
\bauthor{\bsnm{Efimov}, \binits{V.}}:
\batitle{Energy levels arising from resonant two-body forces in a three-body
  system}.
\bjtitle{Physics Letters B}
\bvolume{33}(\bissue{8}),
\bfpage{563}--\blpage{564}
(\byear{1970})
\doiurl{10.1016/0370-2693(70)90349-7}
\end{barticle}
\endbibitem

\bibitem[\protect\citeauthoryear{Efimov}{1971}]{Efimov1971}
\begin{barticle}
\bauthor{\bsnm{Efimov}, \binits{V.N.}}:
\batitle{{Weakly-Bound States of Three Resonantly-interacting Particles}}.
\bjtitle{Soviet Journal of Nuclear Physics}
\bvolume{12}(\bissue{5}),
\bfpage{589}
(\byear{1971})
\end{barticle}
\endbibitem

\bibitem[\protect\citeauthoryear{Efimov}{1973}]{Efimov1973}
\begin{barticle}
\bauthor{\bsnm{Efimov}, \binits{V.}}:
\batitle{Energy levels of three resonantly interacting particles}.
\bjtitle{Nuclear Physics A}
\bvolume{210}(\bissue{1}),
\bfpage{157}--\blpage{188}
(\byear{1973})
\doiurl{10.1016/0375-9474(73)90510-1}
\end{barticle}
\endbibitem

\bibitem[\protect\citeauthoryear{Efimov}{1979}]{Efimov1979}
\begin{barticle}
\bauthor{\bsnm{Efimov}, \binits{V.N.}}:
\batitle{{Low-energy properties of three resonantly interacting particles}}.
\bjtitle{Soviet Journal of Nuclear Physics}
\bvolume{29}(\bissue{4}),
\bfpage{546}
(\byear{1979})
\end{barticle}
\endbibitem

\bibitem[\protect\citeauthoryear{Kraemer et~al.}{2006}]{kraemer2006}
\begin{barticle}
\bauthor{\bsnm{Kraemer}, \binits{T.}},
\bauthor{\bsnm{Mark}, \binits{M.}},
\bauthor{\bsnm{Waldburger}, \binits{P.}},
\bauthor{\bsnm{Danzl}, \binits{J.G.}},
\bauthor{\bsnm{Chin}, \binits{C.}},
\bauthor{\bsnm{Engeser}, \binits{B.}},
\bauthor{\bsnm{Lange}, \binits{A.D.}},
\bauthor{\bsnm{Pilch}, \binits{K.}},
\bauthor{\bsnm{Jaakkola}, \binits{A.}},
\bauthor{\bsnm{N{\"{a}}gerl}, \binits{H.-C.}},
\bauthor{\bsnm{Grimm}, \binits{R.}}:
\batitle{{Evidence for Efimov quantum states in an ultracold gas of caesium
  atoms}}.
\bjtitle{Nature}
\bvolume{440}(\bissue{7082}),
\bfpage{315}--\blpage{318}
(\byear{2006})
\doiurl{10.1038/nature04626}
\end{barticle}
\endbibitem

\bibitem[\protect\citeauthoryear{Ferlaino et~al.}{2011}]{Ferlaino2011}
\begin{barticle}
\bauthor{\bsnm{Ferlaino}, \binits{F.}},
\bauthor{\bsnm{Zenesini}, \binits{A.}},
\bauthor{\bsnm{Berninger}, \binits{M.}},
\bauthor{\bsnm{Huang}, \binits{B.}},
\bauthor{\bsnm{N{\"{a}}gerl}, \binits{H.C.}},
\bauthor{\bsnm{Grimm}, \binits{R.}}:
\batitle{{Efimov Resonances in Ultracold Quantum Gases}}.
\bjtitle{Few-Body Systems}
\bvolume{51}(\bissue{2-4}),
\bfpage{113}--\blpage{133}
(\byear{2011})
\doiurl{10.1007/s00601-011-0260-7}
{\href{https://arxiv.org/abs/1108.1909}{{arXiv:1108.1909}}}
\end{barticle}
\endbibitem

\bibitem[\protect\citeauthoryear{Chin et~al.}{2010}]{Chin2010}
\begin{barticle}
\bauthor{\bsnm{Chin}, \binits{C.}},
\bauthor{\bsnm{Grimm}, \binits{R.}},
\bauthor{\bsnm{Julienne}, \binits{P.}},
\bauthor{\bsnm{Tiesinga}, \binits{E.}}:
\batitle{{Feshbach resonances in ultracold gases}}.
\bjtitle{Reviews of Modern Physics}
\bvolume{82}(\bissue{2}),
\bfpage{1225}--\blpage{1286}
(\byear{2010})
\doiurl{10.1103/RevModPhys.82.1225}
\end{barticle}
\endbibitem

\bibitem[\protect\citeauthoryear{Williams et~al.}{2009}]{Williams2009}
\begin{barticle}
\bauthor{\bsnm{Williams}, \binits{J.R.}},
\bauthor{\bsnm{Hazlett}, \binits{E.L.}},
\bauthor{\bsnm{Huckans}, \binits{J.H.}},
\bauthor{\bsnm{Stites}, \binits{R.W.}},
\bauthor{\bsnm{Zhang}, \binits{Y.}},
\bauthor{\bsnm{O'Hara}, \binits{K.M.}}:
\batitle{{Evidence for an Excited-State Efimov Trimer in a Three-Component
  Fermi Gas}}.
\bjtitle{Physical Review Letters}
\bvolume{103}(\bissue{13}),
\bfpage{130404}
(\byear{2009})
\doiurl{10.1103/PhysRevLett.103.130404}
\end{barticle}
\endbibitem

\bibitem[\protect\citeauthoryear{Zaccanti et~al.}{2009}]{Zaccanti2009}
\begin{barticle}
\bauthor{\bsnm{Zaccanti}, \binits{M.}},
\bauthor{\bsnm{Deissler}, \binits{B.}},
\bauthor{\bsnm{D'Errico}, \binits{C.}},
\bauthor{\bsnm{Fattori}, \binits{M.}},
\bauthor{\bsnm{Jona-Lasinio}, \binits{M.}},
\bauthor{\bsnm{M{\"{u}}ller}, \binits{S.}},
\bauthor{\bsnm{Roati}, \binits{G.}},
\bauthor{\bsnm{Inguscio}, \binits{M.}},
\bauthor{\bsnm{Modugno}, \binits{G.}}:
\batitle{{Observation of an Efimov spectrum in an atomic system}}.
\bjtitle{Nature Physics}
\bvolume{5}(\bissue{8}),
\bfpage{586}--\blpage{591}
(\byear{2009})
\doiurl{10.1038/nphys1334}
\end{barticle}
\endbibitem

\bibitem[\protect\citeauthoryear{Barontini et~al.}{2009}]{Barontini2009}
\begin{barticle}
\bauthor{\bsnm{Barontini}, \binits{G.}},
\bauthor{\bsnm{Weber}, \binits{C.}},
\bauthor{\bsnm{Rabatti}, \binits{F.}},
\bauthor{\bsnm{Catani}, \binits{J.}},
\bauthor{\bsnm{Thalhammer}, \binits{G.}},
\bauthor{\bsnm{Inguscio}, \binits{M.}},
\bauthor{\bsnm{Minardi}, \binits{F.}}:
\batitle{{Observation of Heteronuclear Atomic Efimov Resonances}}.
\bjtitle{Physical Review Letters}
\bvolume{103}(\bissue{4}),
\bfpage{043201}
(\byear{2009})
\doiurl{10.1103/PhysRevLett.103.043201}
\end{barticle}
\endbibitem

\bibitem[\protect\citeauthoryear{Barontini et~al.}{2010}]{Barontini2010}
\begin{barticle}
\bauthor{\bsnm{Barontini}, \binits{G.}},
\bauthor{\bsnm{Weber}, \binits{C.}},
\bauthor{\bsnm{Rabatti}, \binits{F.}},
\bauthor{\bsnm{Catani}, \binits{J.}},
\bauthor{\bsnm{Thalhammer}, \binits{G.}},
\bauthor{\bsnm{Inguscio}, \binits{M.}},
\bauthor{\bsnm{Minardi}, \binits{F.}}:
\batitle{{Erratum: Observation of Heteronuclear Atomic Efimov Resonances [Phys.
  Rev. Lett. 103 , 043201 (2009)]}}.
\bjtitle{Physical Review Letters}
\bvolume{104}(\bissue{5}),
\bfpage{059901}
(\byear{2010})
\doiurl{10.1103/PhysRevLett.104.059901}
\end{barticle}
\endbibitem

\bibitem[\protect\citeauthoryear{Bloom et~al.}{2013}]{Bloom2013a}
\begin{barticle}
\bauthor{\bsnm{Bloom}, \binits{R.S.}},
\bauthor{\bsnm{Hu}, \binits{M.-G.}},
\bauthor{\bsnm{Cumby}, \binits{T.D.}},
\bauthor{\bsnm{Jin}, \binits{D.S.}}:
\batitle{{Tests of Universal Three-Body Physics in an Ultracold Bose-Fermi
  Mixture}}.
\bjtitle{Physical Review Letters}
\bvolume{111}(\bissue{10}),
\bfpage{105301}
(\byear{2013})
\doiurl{10.1103/PhysRevLett.111.105301}
{\href{https://arxiv.org/abs/1304.6989}{{arXiv:1304.6989}}}
\end{barticle}
\endbibitem

\bibitem[\protect\citeauthoryear{Hu et~al.}{2014}]{Hu2014a}
\begin{barticle}
\bauthor{\bsnm{Hu}, \binits{M.-G.}},
\bauthor{\bsnm{Bloom}, \binits{R.S.}},
\bauthor{\bsnm{Jin}, \binits{D.S.}},
\bauthor{\bsnm{Goldwin}, \binits{J.M.}}:
\batitle{{Avalanche-mechanism loss at an atom-molecule Efimov resonance}}.
\bjtitle{Physical Review A}
\bvolume{90}(\bissue{1}),
\bfpage{013619}
(\byear{2014})
\doiurl{10.1103/PhysRevA.90.013619}
\end{barticle}
\endbibitem

\bibitem[\protect\citeauthoryear{Pires et~al.}{2014}]{Pires_2014}
\begin{barticle}
\bauthor{\bsnm{Pires}, \binits{R.}},
\bauthor{\bsnm{Ulmanis}, \binits{J.}},
\bauthor{\bsnm{Häfner}, \binits{S.}},
\bauthor{\bsnm{Repp}, \binits{M.}},
\bauthor{\bsnm{Arias}, \binits{A.}},
\bauthor{\bsnm{Kuhnle}, \binits{E.}},
\bauthor{\bsnm{Weidemüller}, \binits{M.}}:
\batitle{Observation of {E}fimov resonances in a mixture with extreme mass
  imbalance}.
\bjtitle{Phys. Rev. Lett.}
\bvolume{112},
\bfpage{250404}
(\byear{2014})
\doiurl{10.1103/PhysRevLett.112.250404}
\end{barticle}
\endbibitem

\bibitem[\protect\citeauthoryear{Tung et~al.}{2014}]{Tung2014}
\begin{barticle}
\bauthor{\bsnm{Tung}, \binits{S.K.}},
\bauthor{\bsnm{Jim{\'{e}}nez-Garc{\'{i}}a}, \binits{K.}},
\bauthor{\bsnm{Johansen}, \binits{J.}},
\bauthor{\bsnm{Parker}, \binits{C.V.}},
\bauthor{\bsnm{Chin}, \binits{C.}}:
\batitle{{Geometric scaling of efimov states in a ${}^{6}${L}i-${}^{133}${C}s
  mixture}}.
\bjtitle{Physical Review Letters}
\bvolume{113}(\bissue{24}),
\bfpage{1}--\blpage{5}
(\byear{2014})
\doiurl{10.1103/PhysRevLett.113.240402}
\end{barticle}
\endbibitem

\bibitem[\protect\citeauthoryear{Maier et~al.}{2015}]{Maier2015}
\begin{barticle}
\bauthor{\bsnm{Maier}, \binits{R.A.W.}},
\bauthor{\bsnm{Eisele}, \binits{M.}},
\bauthor{\bsnm{Tiemann}, \binits{E.}},
\bauthor{\bsnm{Zimmermann}, \binits{C.}}:
\batitle{{Efimov Resonance and Three-Body Parameter in a Lithium-Rubidium
  Mixture}}.
\bjtitle{Physical Review Letters}
\bvolume{115}(\bissue{4}),
\bfpage{043201}
(\byear{2015})
\doiurl{10.1103/PhysRevLett.115.043201}
\end{barticle}
\endbibitem

\bibitem[\protect\citeauthoryear{Kunitski et~al.}{2015}]{Kunitski2015}
\begin{barticle}
\bauthor{\bsnm{Kunitski}, \binits{M.}},
\bauthor{\bsnm{Zeller}, \binits{S.}},
\bauthor{\bsnm{Voigtsberger}, \binits{J.}},
\bauthor{\bsnm{Kalinin}, \binits{A.}},
\bauthor{\bsnm{Schmidt}, \binits{L.P.H.}},
\bauthor{\bsnm{Sch{\"{o}}ffler}, \binits{M.}},
\bauthor{\bsnm{Czasch}, \binits{A.}},
\bauthor{\bsnm{Sch{\"{o}}llkopf}, \binits{W.}},
\bauthor{\bsnm{Grisenti}, \binits{R.E.}},
\bauthor{\bsnm{Jahnke}, \binits{T.}},
\bauthor{\bsnm{Blume}, \binits{D.}},
\bauthor{\bsnm{D{\"{o}}rner}, \binits{R.}}:
\batitle{{Observation of the Efimov state of the helium trimer}}.
\bjtitle{Journal of Physics: Conference Series}
\bvolume{635}(\bissue{11}),
\bfpage{112096}
(\byear{2015})
\doiurl{10.1088/1742-6596/635/11/112096}
\end{barticle}
\endbibitem

\bibitem[\protect\citeauthoryear{Ulmanis et~al.}{2016}]{Ulmanis2016b}
\begin{barticle}
\bauthor{\bsnm{Ulmanis}, \binits{J.}},
\bauthor{\bsnm{H{\"{a}}fner}, \binits{S.}},
\bauthor{\bsnm{Kuhnle}, \binits{E.D.}},
\bauthor{\bsnm{Weidem{\"{u}}ller}, \binits{M.}}:
\batitle{{Heteronuclear Efimov resonances in ultracold quantum gases}}.
\bjtitle{National Science Review}
\bvolume{3}(\bissue{2}),
\bfpage{174}--\blpage{188}
(\byear{2016})
\doiurl{10.1093/nsr/nww018}
\end{barticle}
\endbibitem

\bibitem[\protect\citeauthoryear{D'Incao and Esry}{2006}]{DIncao2006}
\begin{barticle}
\bauthor{\bsnm{D'Incao}, \binits{J.P.}},
\bauthor{\bsnm{Esry}, \binits{B.D.}}:
\batitle{{E}nhancing the observability of the {E}fimov effect in ultracold
  atomic gas mixtures}.
\bjtitle{Phys. Rev. A}
\bvolume{73},
\bfpage{030703}
(\byear{2006})
\doiurl{10.1103/PhysRevA.73.030703}
\end{barticle}
\endbibitem

\bibitem[\protect\citeauthoryear{Repp et~al.}{2013}]{Repp2013}
\begin{barticle}
\bauthor{\bsnm{Repp}, \binits{M.}},
\bauthor{\bsnm{Pires}, \binits{R.}},
\bauthor{\bsnm{Ulmanis}, \binits{J.}},
\bauthor{\bsnm{Heck}, \binits{R.}},
\bauthor{\bsnm{Kuhnle}, \binits{E.D.}},
\bauthor{\bsnm{Weidem\"uller}, \binits{M.}},
\bauthor{\bsnm{Tiemann}, \binits{E.}}:
\batitle{{O}bservation of interspecies ${}^{6}${L}i-${}^{133}${C}s {F}eshbach
  resonances}.
\bjtitle{Phys. Rev. A}
\bvolume{87},
\bfpage{010701}
(\byear{2013})
\doiurl{10.1103/PhysRevA.87.010701}
\end{barticle}
\endbibitem

\bibitem[\protect\citeauthoryear{Tung et~al.}{2013}]{Tung2013}
\begin{barticle}
\bauthor{\bsnm{Tung}, \binits{S.-K.}},
\bauthor{\bsnm{Parker}, \binits{C.}},
\bauthor{\bsnm{Johansen}, \binits{J.}},
\bauthor{\bsnm{Chin}, \binits{C.}},
\bauthor{\bsnm{Wang}, \binits{Y.}},
\bauthor{\bsnm{Julienne}, \binits{P.S.}}:
\batitle{{U}ltracold mixtures of atomic ${}^{6}${L}i and ${}^{133}${C}s with
  tunable interactions}.
\bjtitle{Phys. Rev. A}
\bvolume{87},
\bfpage{010702}
(\byear{2013})
\doiurl{10.1103/PhysRevA.87.010702}
\end{barticle}
\endbibitem

\bibitem[\protect\citeauthoryear{Ulmanis et~al.}{2016a}]{Ulmanis_2016}
\begin{barticle}
\bauthor{\bsnm{Ulmanis}, \binits{J.}},
\bauthor{\bsnm{Häfner}, \binits{S.}},
\bauthor{\bsnm{Pires}, \binits{R.}},
\bauthor{\bsnm{Kuhnle}, \binits{E.D.}},
\bauthor{\bsnm{Wang}, \binits{Y.}},
\bauthor{\bsnm{Greene}, \binits{C.H.}},
\bauthor{\bsnm{Weidemüller}, \binits{M.}}:
\batitle{Heteronuclear {E}fimov scenario with positive intraspecies scattering
  length}.
\bjtitle{Phys. Rev. Lett.}
\bvolume{117},
\bfpage{153201}
(\byear{2016})
\doiurl{10.1103/PhysRevLett.117.153201}
\end{barticle}
\endbibitem

\bibitem[\protect\citeauthoryear{Ulmanis et~al.}{2016b}]{Ulmanis_2016a}
\begin{barticle}
\bauthor{\bsnm{Ulmanis}, \binits{J.}},
\bauthor{\bsnm{Häfner}, \binits{S.}},
\bauthor{\bsnm{Pires}, \binits{R.}},
\bauthor{\bsnm{Werner}, \binits{F.}},
\bauthor{\bsnm{Petrov}, \binits{D.S.}},
\bauthor{\bsnm{Kuhnle}, \binits{E.D.}},
\bauthor{\bsnm{Weidemüller}, \binits{M.}}:
\batitle{Universal three-body recombination and {E}fimov resonances in an
  ultracold {L}i-{C}s mixture}.
\bjtitle{Phys. Rev. A}
\bvolume{93}(\bissue{2}),
\bfpage{022707}
(\byear{2016})
\doiurl{10.1103/PhysRevA.93.022707}
\end{barticle}
\endbibitem

\bibitem[\protect\citeauthoryear{Ulmanis}{2015}]{Ulmanis2015}
\begin{botherref}
\oauthor{\bsnm{Ulmanis}, \binits{J.}}:
{Universality and non-universality in the heteronuclear Efimov scenario with
  large mass imbalance}.
PhD thesis,
Ruprecht-Karls-Universit{\"{a}}t Heidelberg
(2015).
\doiurl{10.11588/heidok.00019214}
\end{botherref}
\endbibitem

\bibitem[\protect\citeauthoryear{Weiner et~al.}{1999}]{Weiner1999}
\begin{barticle}
\bauthor{\bsnm{Weiner}, \binits{J.}},
\bauthor{\bsnm{Bagnato}, \binits{V.S.}},
\bauthor{\bsnm{Zilio}, \binits{S.}},
\bauthor{\bsnm{Julienne}, \binits{P.S.}}:
\batitle{{Experiments and theory in cold and ultracold collisions}}.
\bjtitle{Reviews of Modern Physics}
\bvolume{71}(\bissue{1}),
\bfpage{1}--\blpage{85}
(\byear{1999})
\doiurl{10.1103/RevModPhys.71.1}
\end{barticle}
\endbibitem

\bibitem[\protect\citeauthoryear{Sch{\"{u}}nemann
  et~al.}{1998}]{Schunemann1998}
\begin{barticle}
\bauthor{\bsnm{Sch{\"{u}}nemann}, \binits{U.}},
\bauthor{\bsnm{Engler}, \binits{H.}},
\bauthor{\bsnm{Zielonkowski}, \binits{M.}},
\bauthor{\bsnm{Weidem{\"{u}}ller}, \binits{M.}},
\bauthor{\bsnm{Grimm}, \binits{R.}}:
\batitle{{Magneto-optic trapping of lithium using semiconductor lasers}}.
\bjtitle{Optics Communications}
\bvolume{158}(\bissue{1-6}),
\bfpage{263}--\blpage{272}
(\byear{1998})
\doiurl{10.1016/S0030-4018(98)00517-3}
\end{barticle}
\endbibitem

\bibitem[\protect\citeauthoryear{Metcalf and van~der
  Straten}{1999}]{MetcalfStraten1999}
\begin{bbook}
\bauthor{\bsnm{Metcalf}, \binits{H.J.}},
\bauthor{\bsnm{Straten}, \binits{P.}}:
\bbtitle{{Laser Cooling and Trapping}}.
\bpublisher{Springer},
\blocation{New York}
(\byear{1999})
\end{bbook}
\endbibitem

\bibitem[\protect\citeauthoryear{Cohen-Tannoudji et~al.}{2011}]{Cohen2011}
\begin{bbook}
\bauthor{\bsnm{Cohen-Tannoudji}, \binits{C.}}, \betal:
\bbtitle{Advances in Atomic Physics: an Overview}.
\bpublisher{World scientific},
\blocation{Hackensack, NJ}
(\byear{2011})
\end{bbook}
\endbibitem

\bibitem[\protect\citeauthoryear{Ketterle and Druten}{1996}]{Ketterle1996}
\begin{botherref}
\oauthor{\bsnm{Ketterle}, \binits{W.}},
\oauthor{\bsnm{Druten}, \binits{N.J.V.}}:
Evaporative cooling of trapped atoms.
Advances In Atomic, Molecular, and Optical Physics,
vol. 37,
pp. 181--236.
Academic Press
(1996).
\doiurl{10.1016/S1049-250X(08)60101-9}
\end{botherref}
\endbibitem

\bibitem[\protect\citeauthoryear{Grimm et~al.}{2000}]{Grimm2000}
\begin{botherref}
\oauthor{\bsnm{Grimm}, \binits{R.}},
\oauthor{\bsnm{Weidem{\"{u}}ller}, \binits{M.}},
\oauthor{\bsnm{Ovchinnikov}, \binits{Y.B.}}:
Optical dipole traps for neutral atoms.
Advances In Atomic, Molecular, and Optical Physics,
vol. 42,
pp. 95--170.
Academic Press
(2000).
\doiurl{10.1016/S1049-250X(08)60186-X}
\end{botherref}
\endbibitem

\bibitem[\protect\citeauthoryear{Phillips and Metcalf}{1982}]{Phillips1982}
\begin{barticle}
\bauthor{\bsnm{Phillips}, \binits{W.D.}},
\bauthor{\bsnm{Metcalf}, \binits{H.}}:
\batitle{{Laser Deceleration of an Atomic Beam}}.
\bjtitle{Physical Review Letters}
\bvolume{48}(\bissue{9}),
\bfpage{596}--\blpage{599}
(\byear{1982})
\doiurl{10.1103/PhysRevLett.48.596}
\end{barticle}
\endbibitem

\bibitem[\protect\citeauthoryear{Häfner et~al.}{2017}]{Haefner_2017}
\begin{barticle}
\bauthor{\bsnm{Häfner}, \binits{S.}},
\bauthor{\bsnm{Ulmanis}, \binits{J.}},
\bauthor{\bsnm{Kuhnle}, \binits{E.D.}},
\bauthor{\bsnm{Wang}, \binits{Y.}},
\bauthor{\bsnm{Greene}, \binits{C.H.}},
\bauthor{\bsnm{Weidemüller}, \binits{M.}}:
\batitle{Role of the intraspecies scattering length in the efimov scenario with
  large mass difference}.
\bjtitle{Phys. Rev. A}
\bvolume{95},
\bfpage{062708}
(\byear{2017})
\doiurl{10.1103/PhysRevA.95.062708}
\end{barticle}
\endbibitem

\bibitem[\protect\citeauthoryear{LeBlanc and Thywissen}{2007}]{LeBlanc2007}
\begin{barticle}
\bauthor{\bsnm{LeBlanc}, \binits{L.J.}},
\bauthor{\bsnm{Thywissen}, \binits{J.H.}}:
\batitle{Species-specific optical lattices}.
\bjtitle{Phys. Rev. A}
\bvolume{75},
\bfpage{053612}
(\byear{2007})
\doiurl{10.1103/PhysRevA.75.053612}
\end{barticle}
\endbibitem

\bibitem[\protect\citeauthoryear{Repp}{2013}]{Repp2013PhD}
\begin{botherref}
\oauthor{\bsnm{Repp}, \binits{M.}}:
{Interspecies Feshbach Resonances in an ultracold, optically trapped Bose-Fermi
  mixture of Cesium an Lithium}.
PhD thesis,
Ruprecht-Karls-Universit{\"{a}}t Heidelberg
(2013).
\doiurl{10.11588/heidok.00014993}
\end{botherref}
\endbibitem

\bibitem[\protect\citeauthoryear{Stan and Ketterle}{2005}]{Stan2005}
\begin{barticle}
\bauthor{\bsnm{Stan}, \binits{C.A.}},
\bauthor{\bsnm{Ketterle}, \binits{W.}}:
\batitle{{M}ultiple species atom source for laser-cooling experiments}.
\bjtitle{Review of Scientific Instruments}
\bvolume{76}(\bissue{6}),
\bfpage{063113}
(\byear{2005})
\doiurl{10.1063/1.1935433}
\end{barticle}
\endbibitem

\bibitem[\protect\citeauthoryear{Gehm}{2016}]{Gehm2003}
\begin{botherref}
\oauthor{\bsnm{Gehm}, \binits{M.E.}}:
\textsc{P}roperties of $^6{L}i$.
available online at
  https://jet.physics.ncsu.edu/techdocs/pdf/PropertiesOfLi.pdf (version updated
  11/2016)
(2016).
\url{https://jet.physics.ncsu.edu/techdocs}
\end{botherref}
\endbibitem

\bibitem[\protect\citeauthoryear{Steck}{2024}]{Steck2008}
\begin{botherref}
\oauthor{\bsnm{Steck}, \binits{D.A.}}:
{C}esium {D} {L}ine {D}ata.
available online at http://steck.us/alkalidata (Version 2.3.3, last revised 28
  May 2024)
(2024).
\url{http://steck.us/alkalidata}
\end{botherref}
\endbibitem

\bibitem[\protect\citeauthoryear{Bell et~al.}{2010}]{Bell2010}
\begin{barticle}
\bauthor{\bsnm{Bell}, \binits{S.C.}},
\bauthor{\bsnm{Junker}, \binits{M.}},
\bauthor{\bsnm{Jasperse}, \binits{M.}},
\bauthor{\bsnm{Turner}, \binits{L.D.}},
\bauthor{\bsnm{Lin}, \binits{Y.-J.}},
\bauthor{\bsnm{Spielman}, \binits{I.B.}},
\bauthor{\bsnm{Scholten}, \binits{R.E.}}:
\batitle{{A} slow atom source using a collimated effusive oven and a
  single-layer variable pitch coil {Z}eeman slower}.
\bjtitle{Review of Scientific Instruments}
\bvolume{81}(\bissue{1}),
\bfpage{013105}
(\byear{2010})
\doiurl{10.1063/1.3276712}
\end{barticle}
\endbibitem

\bibitem[\protect\citeauthoryear{Joffe et~al.}{1993}]{Joffe1993}
\begin{barticle}
\bauthor{\bsnm{Joffe}, \binits{M.A.}},
\bauthor{\bsnm{Ketterle}, \binits{W.}},
\bauthor{\bsnm{Martin}, \binits{A.}},
\bauthor{\bsnm{Pritchard}, \binits{D.E.}}:
\batitle{{T}ransverse cooling and deflection of an atomic beam inside a
  {Z}eeman slower}.
\bjtitle{J. Opt. Soc. Am. B}
\bvolume{10}(\bissue{12}),
\bfpage{2257}--\blpage{2262}
(\byear{1993})
\doiurl{10.1364/JOSAB.10.002257}
\end{barticle}
\endbibitem

\bibitem[\protect\citeauthoryear{Marti et~al.}{2010}]{Marti2010}
\begin{barticle}
\bauthor{\bsnm{Marti}, \binits{G.E.}},
\bauthor{\bsnm{Olf}, \binits{R.}},
\bauthor{\bsnm{Vogt}, \binits{E.}},
\bauthor{\bsnm{\"Ottl}, \binits{A.}},
\bauthor{\bsnm{Stamper-Kurn}, \binits{D.M.}}:
\batitle{{T}wo-element {Z}eeman slower for rubidium and lithium}.
\bjtitle{Phys. Rev. A}
\bvolume{81},
\bfpage{043424}
(\byear{2010})
\doiurl{10.1103/PhysRevA.81.043424}
\end{barticle}
\endbibitem

\bibitem[\protect\citeauthoryear{Weber et~al.}{2003}]{Weber2003a}
\begin{barticle}
\bauthor{\bsnm{Weber}, \binits{T.}},
\bauthor{\bsnm{Herbig}, \binits{J.}},
\bauthor{\bsnm{Mark}, \binits{M.}},
\bauthor{\bsnm{Naegerl}, \binits{H.-C.}},
\bauthor{\bsnm{Grimm}, \binits{R.}}:
\batitle{{Bose-Einstein Condensation of Cesium}}.
\bjtitle{Science}
\bvolume{299}(\bissue{5604}),
\bfpage{232}--\blpage{235}
(\byear{2003})
\doiurl{10.1126/science.1079699}
\end{barticle}
\endbibitem

\bibitem[\protect\citeauthoryear{Lippi}{2024}]{Lippi2024}
\begin{botherref}
\oauthor{\bsnm{Lippi}, \binits{E.}}:
{${}^{133}${C}s atoms in a ${}^{6}${L}i Fermi sea for exploring polaron physics
  in the heavy impurity limit}.
PhD thesis,
Ruprecht-Karls-Universit{\"{a}}t Heidelberg
(2024).
\doiurl{10.11588/heidok.00035527}
\end{botherref}
\endbibitem

\bibitem[\protect\citeauthoryear{Tran}{2022}]{Tran2022}
\begin{botherref}
\oauthor{\bsnm{Tran}, \binits{B.}}:
{From Efimov Physics to Polarons in an Ultracold Mixture of Li and Cs Atoms}.
PhD thesis,
Ruprecht-Karls-Universit{\"{a}}t Heidelberg
(2022).
\doiurl{10.11588/heidok.00031882}
\end{botherref}
\endbibitem

\bibitem[\protect\citeauthoryear{Schünemann et~al.}{1998}]{Shunemann1999}
\begin{barticle}
\bauthor{\bsnm{Schünemann}, \binits{U.}},
\bauthor{\bsnm{Engler}, \binits{H.}},
\bauthor{\bsnm{Zielonkowski}, \binits{M.}},
\bauthor{\bsnm{Weidemüller}, \binits{M.}},
\bauthor{\bsnm{Grimm}, \binits{R.}}:
\batitle{Magneto-optic trapping of lithium using semiconductor lasers}.
\bjtitle{Optics Communications}
\bvolume{158}(\bissue{1}),
\bfpage{263}--\blpage{272}
(\byear{1998})
\doiurl{10.1016/S0030-4018(98)00517-3}
\end{barticle}
\endbibitem

\bibitem[\protect\citeauthoryear{Schl{\"{o}}der et~al.}{1999}]{Schloder1999}
\begin{barticle}
\bauthor{\bsnm{Schl{\"{o}}der}, \binits{U.}},
\bauthor{\bsnm{Engler}, \binits{H.}},
\bauthor{\bsnm{Sch{\"{u}}nemann}, \binits{U.}},
\bauthor{\bsnm{Grimm}, \binits{R.}},
\bauthor{\bsnm{Weidem{\"{u}}ller}, \binits{M.}}:
\batitle{{Cold inelastic collisions between lithium and cesium in a two-species
  magneto-optical trap}}.
\bjtitle{The European Physical Journal D}
\bvolume{7}(\bissue{3}),
\bfpage{331}
(\byear{1999})
\doiurl{10.1007/s100530050576}
\end{barticle}
\endbibitem

\bibitem[\protect\citeauthoryear{Repp et~al.}{2013}]{Repp_2013}
\begin{barticle}
\bauthor{\bsnm{Repp}, \binits{M.}},
\bauthor{\bsnm{Pires}, \binits{R.}},
\bauthor{\bsnm{Ulmanis}, \binits{J.}},
\bauthor{\bsnm{Heck}, \binits{R.}},
\bauthor{\bsnm{Kuhnle}, \binits{E.D.}},
\bauthor{\bsnm{Weidemüller}, \binits{M.}},
\bauthor{\bsnm{Tiemann}, \binits{E.}}:
\batitle{Observation of interspecies ${}^{6}${L}i-${}^{133}${C}s {F}eshbach
  resonances}.
\bjtitle{Phys. Rev. A}
\bvolume{87},
\bfpage{010701}
(\byear{2013})
\doiurl{10.1103/PhysRevA.87.010701}
\end{barticle}
\endbibitem

\bibitem[\protect\citeauthoryear{Zhu et~al.}{2019a}]{zhu2019}
\begin{botherref}
\oauthor{\bsnm{Zhu}, \binits{B.}},
\oauthor{\bsnm{H{\"{a}}fner}, \binits{S.}},
\oauthor{\bsnm{Tran}, \binits{B.}},
\oauthor{\bsnm{Gerken}, \binits{M.}},
\oauthor{\bsnm{Ulmanis}, \binits{J.}},
\oauthor{\bsnm{Tiemann}, \binits{E.}},
\oauthor{\bsnm{Weidem{\"{u}}ller}, \binits{M.}}:
{Spin-rotation coupling in p-wave Feshbach resonances}
(2019)
{\href{https://arxiv.org/abs/1910.12011}{{arXiv:1910.12011}}}
\end{botherref}
\endbibitem

\bibitem[\protect\citeauthoryear{Zhu et~al.}{2019b}]{zhu2019a}
\begin{botherref}
\oauthor{\bsnm{Zhu}, \binits{B.}},
\oauthor{\bsnm{H{\"{a}}fner}, \binits{S.}},
\oauthor{\bsnm{Tran}, \binits{B.}},
\oauthor{\bsnm{Gerken}, \binits{M.}},
\oauthor{\bsnm{Ulmanis}, \binits{J.}},
\oauthor{\bsnm{Tiemann}, \binits{E.}},
\oauthor{\bsnm{Weidem{\"{u}}ller}, \binits{M.}}:
{High partial-wave Feshbach resonances in an ultracold
  ${}^{6}${L}i-${}^{133}${C}s mixture}
(2019)
{\href{https://arxiv.org/abs/1912.01264}{{arXiv:1912.01264}}}
\end{botherref}
\endbibitem

\bibitem[\protect\citeauthoryear{Drewsen et~al.}{1994}]{Drewsen1994}
\begin{barticle}
\bauthor{\bsnm{Drewsen}, \binits{M.}},
\bauthor{\bsnm{Laurent}, \binits{P.}},
\bauthor{\bsnm{Nadir}, \binits{A.}},
\bauthor{\bsnm{Santarelli}, \binits{G.}},
\bauthor{\bsnm{Clairon}, \binits{A.}},
\bauthor{\bsnm{Castin}, \binits{Y.}},
\bauthor{\bsnm{Grison}, \binits{D.}},
\bauthor{\bsnm{Salomon}, \binits{C.}}:
\batitle{{Investigation of sub-Doppler cooling effects in a cesium
  magneto-optical trap}}.
\bjtitle{Applied Physics B}
\bvolume{59}(\bissue{3}),
\bfpage{283}--\blpage{298}
(\byear{1994})
\doiurl{10.1007/BF01081396}
\end{barticle}
\endbibitem

\bibitem[\protect\citeauthoryear{Vuleti\'{c} et~al.}{1998}]{Vuletic1998}
\begin{barticle}
\bauthor{\bsnm{Vuleti\'{c}}, \binits{V.}},
\bauthor{\bsnm{Chin}, \binits{C.}},
\bauthor{\bsnm{Kerman}, \binits{A.J.}},
\bauthor{\bsnm{Chu}, \binits{S.}}:
\batitle{Degenerate {R}aman sideband cooling of trapped cesium atoms at very
  high atomic densities}.
\bjtitle{Phys. Rev. Lett.}
\bvolume{81},
\bfpage{5768}--\blpage{5771}
(\byear{1998})
\doiurl{10.1103/PhysRevLett.81.5768}
\end{barticle}
\endbibitem

\bibitem[\protect\citeauthoryear{Kerman et~al.}{2000}]{Kerman2000}
\begin{barticle}
\bauthor{\bsnm{Kerman}, \binits{A.J.}},
\bauthor{\bsnm{Vuletic}, \binits{V.}},
\bauthor{\bsnm{Chin}, \binits{C.}},
\bauthor{\bsnm{Chu}, \binits{S.}}:
\batitle{{B}eyond {O}ptical {M}olasses: 3{D} {R}aman {S}ideband {C}ooling of
  {A}tomic {C}esium to {H}igh {P}hase-{S}pace {D}ensity}.
\bjtitle{Phys. Rev. Lett.}
\bvolume{84},
\bfpage{439}--\blpage{442}
(\byear{2000})
\doiurl{10.1103/PhysRevLett.84.439}
\end{barticle}
\endbibitem

\bibitem[\protect\citeauthoryear{Treutlein et~al.}{2001}]{Treutlein2001}
\begin{barticle}
\bauthor{\bsnm{Treutlein}, \binits{P.}},
\bauthor{\bsnm{Chung}, \binits{K.Y.}},
\bauthor{\bsnm{Chu}, \binits{S.}}:
\batitle{High-brightness atom source for atomic fountains}.
\bjtitle{Phys. Rev. A}
\bvolume{63},
\bfpage{051401}
(\byear{2001})
\doiurl{10.1103/PhysRevA.63.051401}
\end{barticle}
\endbibitem

\bibitem[\protect\citeauthoryear{Petrov and Werner}{2015}]{Petrov_2015}
\begin{botherref}
\oauthor{\bsnm{Petrov}, \binits{D.S.}},
\oauthor{\bsnm{Werner}, \binits{F.}}:
Three-body recombination in heteronuclear mixtures at finite temperature.
Physical Review A
\textbf{92}(2)
(2015)
\doiurl{10.1103/physreva.92.022704}
\end{botherref}
\endbibitem

\bibitem[\protect\citeauthoryear{Gerken et~al.}{2019}]{Gerken2019}
\begin{barticle}
\bauthor{\bsnm{Gerken}, \binits{M.}},
\bauthor{\bsnm{Tran}, \binits{B.}},
\bauthor{\bsnm{H\"afner}, \binits{S.}},
\bauthor{\bsnm{Tiemann}, \binits{E.}},
\bauthor{\bsnm{Zhu}, \binits{B.}},
\bauthor{\bsnm{Weidem\"uller}, \binits{M.}}:
\batitle{Observation of dipolar splittings in high-resolution atom-loss
  spectroscopy of $^{6}\mathrm{Li}$ $p$-wave feshbach resonances}.
\bjtitle{Phys. Rev. A}
\bvolume{100},
\bfpage{050701}
(\byear{2019})
\doiurl{10.1103/PhysRevA.100.050701}
\end{barticle}
\endbibitem

\bibitem[\protect\citeauthoryear{Gerken}{2022}]{Gerken2022}
\begin{botherref}
\oauthor{\bsnm{Gerken}, \binits{M.}}:
{Exploring p-wave Feshbach Resonances in Ultracold Lithium and Lithium-Cesium
  Mixtures}.
PhD thesis,
Ruprecht-Karls-Universit{\"{a}}t Heidelberg
(2022).
\doiurl{10.11588/heidok.00031719}
\end{botherref}
\endbibitem

\bibitem[\protect\citeauthoryear{H{\"{a}}fner}{2017}]{Hafner2017a}
\begin{botherref}
\oauthor{\bsnm{H{\"{a}}fner}, \binits{S.}}:
{From two-body to many-body physics in an ultracold Bose-Fermi mixture of Li
  and Cs atoms}.
PhD thesis,
Ruprecht-Karls-Universit{\"{a}}t Heidelberg,
Heidelberg
(2017).
\doiurl{10.11588/heidok.00023852}
\end{botherref}
\endbibitem

\bibitem[\protect\citeauthoryear{Silber et~al.}{2005}]{Silber2005}
\begin{barticle}
\bauthor{\bsnm{Silber}, \binits{C.}},
\bauthor{\bsnm{G\"unther}, \binits{S.}},
\bauthor{\bsnm{Marzok}, \binits{C.}},
\bauthor{\bsnm{Deh}, \binits{B.}},
\bauthor{\bsnm{Courteille}, \binits{P.W.}},
\bauthor{\bsnm{Zimmermann}, \binits{C.}}:
\batitle{{Q}uantum-{D}egenerate {M}ixture of {F}ermionic {L}ithium and
  {B}osonic {R}ubidium {G}ases}.
\bjtitle{Phys. Rev. Lett.}
\bvolume{95},
\bfpage{170408}
(\byear{2005})
\doiurl{10.1103/PhysRevLett.95.170408}
\end{barticle}
\endbibitem

\bibitem[\protect\citeauthoryear{Hansen et~al.}{2011}]{Hansen2011}
\begin{barticle}
\bauthor{\bsnm{Hansen}, \binits{A.H.}},
\bauthor{\bsnm{Khramov}, \binits{A.}},
\bauthor{\bsnm{Dowd}, \binits{W.H.}},
\bauthor{\bsnm{Jamison}, \binits{A.O.}},
\bauthor{\bsnm{Ivanov}, \binits{V.V.}},
\bauthor{\bsnm{Gupta}, \binits{S.}}:
\batitle{{Q}uantum degenerate mixture of ytterbium and lithium atoms}.
\bjtitle{Phys. Rev. A}
\bvolume{84},
\bfpage{011606}
(\byear{2011})
\doiurl{10.1103/PhysRevA.84.011606}
\end{barticle}
\endbibitem

\bibitem[\protect\citeauthoryear{DeSalvo et~al.}{2017}]{DeSalvo2017}
\begin{barticle}
\bauthor{\bsnm{DeSalvo}, \binits{B.J.}},
\bauthor{\bsnm{Patel}, \binits{K.}},
\bauthor{\bsnm{Johansen}, \binits{J.}},
\bauthor{\bsnm{Chin}, \binits{C.}}:
\batitle{Observation of a degenerate fermi gas trapped by a bose-einstein
  condensate}.
\bjtitle{Phys. Rev. Lett.}
\bvolume{119},
\bfpage{233401}
(\byear{2017})
\doiurl{10.1103/PhysRevLett.119.233401}
\end{barticle}
\endbibitem

\bibitem[\protect\citeauthoryear{Chen et~al.}{2023}]{Chen2023}
\begin{barticle}
\bauthor{\bsnm{Chen}, \binits{Y.}},
\bauthor{\bsnm{Li}, \binits{W.}},
\bauthor{\bsnm{Sun}, \binits{Y.}},
\bauthor{\bsnm{Chen}, \binits{Q.}},
\bauthor{\bsnm{Chang}, \binits{P.}},
\bauthor{\bsnm{Tung}, \binits{S.}}:
\batitle{{Dual-species Bose-Einstein condensates of ${}^{7}${L}i and
  ${}^{133}${C}s}}.
\bjtitle{Physical Review A}
\bvolume{108}(\bissue{033301}),
\bfpage{1}--\blpage{8}
(\byear{2023})
\doiurl{10.1103/PhysRevA.108.033301}
\end{barticle}
\endbibitem

\bibitem[\protect\citeauthoryear{Levinsen et~al.}{2015}]{Levinsen2015}
\begin{barticle}
\bauthor{\bsnm{Levinsen}, \binits{J.}},
\bauthor{\bsnm{Parish}, \binits{M.M.}},
\bauthor{\bsnm{Bruun}, \binits{G.M.}}:
\batitle{{Impurity in a Bose-Einstein Condensate and the Efimov Effect}}.
\bjtitle{Physical Review Letters}
\bvolume{115}(\bissue{12}),
\bfpage{125302}
(\byear{2015})
\doiurl{10.1103/PhysRevLett.115.125302}
\end{barticle}
\endbibitem

\bibitem[\protect\citeauthoryear{Sun et~al.}{2017}]{Sun2017}
\begin{barticle}
\bauthor{\bsnm{Sun}, \binits{M.}},
\bauthor{\bsnm{Zhai}, \binits{H.}},
\bauthor{\bsnm{Cui}, \binits{X.}}:
\batitle{{Visualizing the Efimov Correlation in Bose Polarons}}.
\bjtitle{Physical Review Letters}
\bvolume{119}(\bissue{1}),
\bfpage{013401}
(\byear{2017})
\doiurl{10.1103/PhysRevLett.119.013401}
\end{barticle}
\endbibitem

\bibitem[\protect\citeauthoryear{Sun and Cui}{2017}]{Sun2017a}
\begin{barticle}
\bauthor{\bsnm{Sun}, \binits{M.}},
\bauthor{\bsnm{Cui}, \binits{X.}}:
\batitle{{Enhancing the Efimov correlation in Bose polarons with large mass
  imbalance}}.
\bjtitle{Physical Review A}
\bvolume{96}(\bissue{2}),
\bfpage{022707}
(\byear{2017})
\doiurl{10.1103/PhysRevA.96.022707}
\end{barticle}
\endbibitem

\bibitem[\protect\citeauthoryear{Arora et~al.}{2011}]{Arora2011}
\begin{barticle}
\bauthor{\bsnm{Arora}, \binits{B.}},
\bauthor{\bsnm{Safronova}, \binits{M.S.}},
\bauthor{\bsnm{Clark}, \binits{C.W.}}:
\batitle{{Tune-out wavelengths of alkali-metal atoms and their applications}}.
\bjtitle{Physical Review A}
\bvolume{84}(\bissue{4}),
\bfpage{043401}
(\byear{2011})
\doiurl{10.1103/PhysRevA.84.043401}
\end{barticle}
\endbibitem

\bibitem[\protect\citeauthoryear{Ratkata et~al.}{2021}]{Ratkata2021}
\begin{barticle}
\bauthor{\bsnm{Ratkata}, \binits{A.}},
\bauthor{\bsnm{Gregory}, \binits{P.D.}},
\bauthor{\bsnm{Innes}, \binits{A.D.}},
\bauthor{\bsnm{Matthies}, \binits{A.J.}},
\bauthor{\bsnm{McArd}, \binits{L.A.}},
\bauthor{\bsnm{Mortlock}, \binits{J.M.}},
\bauthor{\bsnm{Safronova}, \binits{M.S.}},
\bauthor{\bsnm{Bromley}, \binits{S.L.}},
\bauthor{\bsnm{Cornish}, \binits{S.L.}}:
\batitle{{Measurement of the tune-out wavelength for ${}^{133}${C}s at 880
  nm}}.
\bjtitle{Physical Review A}
\bvolume{104}(\bissue{5}),
\bfpage{052813}
(\byear{2021})
\doiurl{10.1103/PhysRevA.104.052813}
\end{barticle}
\endbibitem

\bibitem[\protect\citeauthoryear{Hu et~al.}{2016}]{Hu2016}
\begin{barticle}
\bauthor{\bsnm{Hu}, \binits{M.-G.}},
\bauthor{\bsnm{{Van de Graaff}}, \binits{M.J.}},
\bauthor{\bsnm{Kedar}, \binits{D.}},
\bauthor{\bsnm{Corson}, \binits{J.P.}},
\bauthor{\bsnm{Cornell}, \binits{E.A.}},
\bauthor{\bsnm{Jin}, \binits{D.S.}}:
\batitle{{Bose Polarons in the Strongly Interacting Regime}}.
\bjtitle{Physical Review Letters}
\bvolume{117}(\bissue{5}),
\bfpage{055301}
(\byear{2016})
\doiurl{10.1103/PhysRevLett.117.055301}
\end{barticle}
\endbibitem

\bibitem[\protect\citeauthoryear{Kohstall et~al.}{2012}]{Kohstall2012}
\begin{barticle}
\bauthor{\bsnm{Kohstall}, \binits{C.}},
\bauthor{\bsnm{Zaccanti}, \binits{M.}},
\bauthor{\bsnm{Jag}, \binits{M.}},
\bauthor{\bsnm{Trenkwalder}, \binits{A.}},
\bauthor{\bsnm{Massignan}, \binits{P.}},
\bauthor{\bsnm{Bruun}, \binits{G.M.}},
\bauthor{\bsnm{Schreck}, \binits{F.}},
\bauthor{\bsnm{Grimm}, \binits{R.}}:
\batitle{{Metastability and coherence of repulsive polarons in a strongly
  interacting Fermi mixture}}.
\bjtitle{Nature}
\bvolume{485}(\bissue{7400}),
\bfpage{615}--\blpage{618}
(\byear{2012})
\doiurl{10.1038/nature11065}
\end{barticle}
\endbibitem

\bibitem[\protect\citeauthoryear{Cetina et~al.}{2015}]{Cetina2015}
\begin{barticle}
\bauthor{\bsnm{Cetina}, \binits{M.}},
\bauthor{\bsnm{Jag}, \binits{M.}},
\bauthor{\bsnm{Lous}, \binits{R.S.}},
\bauthor{\bsnm{Walraven}, \binits{J.T.M.}},
\bauthor{\bsnm{Grimm}, \binits{R.}},
\bauthor{\bsnm{Christensen}, \binits{R.S.}},
\bauthor{\bsnm{Bruun}, \binits{G.M.}}:
\batitle{{Decoherence of Impurities in a Fermi Sea of Ultracold Atoms}}.
\bjtitle{Physical Review Letters}
\bvolume{115}(\bissue{13}),
\bfpage{135302}
(\byear{2015})
\doiurl{10.1103/PhysRevLett.115.135302}
\end{barticle}
\endbibitem

\bibitem[\protect\citeauthoryear{Cetina et~al.}{2016}]{Cetina2016}
\begin{barticle}
\bauthor{\bsnm{Cetina}, \binits{M.}},
\bauthor{\bsnm{Jag}, \binits{M.}},
\bauthor{\bsnm{Lous}, \binits{R.S.}},
\bauthor{\bsnm{Fritsche}, \binits{I.}},
\bauthor{\bsnm{Walraven}, \binits{J.T.M.}},
\bauthor{\bsnm{Grimm}, \binits{R.}},
\bauthor{\bsnm{Levinsen}, \binits{J.}},
\bauthor{\bsnm{Parish}, \binits{M.M.}},
\bauthor{\bsnm{Schmidt}, \binits{R.}},
\bauthor{\bsnm{Knap}, \binits{M.}},
\bauthor{\bsnm{Demler}, \binits{E.}}:
\batitle{{Ultrafast many-body interferometry of impurities coupled to a Fermi
  sea}}.
\bjtitle{Science}
\bvolume{354}(\bissue{6308}),
\bfpage{96}--\blpage{99}
(\byear{2016})
\doiurl{10.1126/science.aaf5134}
\end{barticle}
\endbibitem

\bibitem[\protect\citeauthoryear{Yan et~al.}{2020}]{Yan2020a}
\begin{barticle}
\bauthor{\bsnm{Yan}, \binits{Z.Z.}},
\bauthor{\bsnm{Ni}, \binits{Y.}},
\bauthor{\bsnm{Robens}, \binits{C.}},
\bauthor{\bsnm{Zwierlein}, \binits{M.W.}}:
\batitle{{Bose polarons near quantum criticality}}.
\bjtitle{Science}
\bvolume{368}(\bissue{6487}),
\bfpage{190}--\blpage{194}
(\byear{2020})
\doiurl{10.1126/science.aax5850}
\end{barticle}
\endbibitem

\bibitem[\protect\citeauthoryear{Fritsche et~al.}{2021}]{Fritsche2021a}
\begin{barticle}
\bauthor{\bsnm{Fritsche}, \binits{I.}},
\bauthor{\bsnm{Baroni}, \binits{C.}},
\bauthor{\bsnm{Dobler}, \binits{E.}},
\bauthor{\bsnm{Kirilov}, \binits{E.}},
\bauthor{\bsnm{Huang}, \binits{B.}},
\bauthor{\bsnm{Grimm}, \binits{R.}},
\bauthor{\bsnm{Bruun}, \binits{G.M.}},
\bauthor{\bsnm{Massignan}, \binits{P.}}:
\batitle{{Stability and breakdown of Fermi polarons in a strongly interacting
  Fermi-Bose mixture}}.
\bjtitle{Physical Review A}
\bvolume{103}(\bissue{5}),
\bfpage{053314}
(\byear{2021})
\doiurl{10.1103/PhysRevA.103.053314}
\end{barticle}
\endbibitem

\bibitem[\protect\citeauthoryear{Baroni et~al.}{2024}]{Baroni2023}
\begin{barticle}
\bauthor{\bsnm{Baroni}, \binits{C.}},
\bauthor{\bsnm{Huang}, \binits{B.}},
\bauthor{\bsnm{Fritsche}, \binits{I.}},
\bauthor{\bsnm{Dobler}, \binits{E.}},
\bauthor{\bsnm{Anich}, \binits{G.}},
\bauthor{\bsnm{Kirilov}, \binits{E.}},
\bauthor{\bsnm{Grimm}, \binits{R.}},
\bauthor{\bsnm{Bastarrachea-Magnani}, \binits{M.A.}},
\bauthor{\bsnm{Massignan}, \binits{P.}},
\bauthor{\bsnm{Bruun}, \binits{G.M.}}:
\batitle{{Mediated interactions between Fermi polarons and the role of impurity
  quantum statistics}}.
\bjtitle{Nature Physics}
\bvolume{20}(\bissue{1}),
\bfpage{68}--\blpage{73}
(\byear{2024})
\doiurl{10.1038/s41567-023-02248-4}
\end{barticle}
\endbibitem

\bibitem[\protect\citeauthoryear{Duda et~al.}{2023}]{Duda2023}
\begin{barticle}
\bauthor{\bsnm{Duda}, \binits{M.}},
\bauthor{\bsnm{Chen}, \binits{X.-Y.}},
\bauthor{\bsnm{Schindewolf}, \binits{A.}},
\bauthor{\bsnm{Bause}, \binits{R.}},
\bauthor{\bsnm{Milczewski}, \binits{J.}},
\bauthor{\bsnm{Schmidt}, \binits{R.}},
\bauthor{\bsnm{Bloch}, \binits{I.}},
\bauthor{\bsnm{Luo}, \binits{X.-Y.}}:
\batitle{{Transition from a polaronic condensate to a degenerate Fermi gas of
  heteronuclear molecules}}.
\bjtitle{Nature Physics}
\bvolume{19}(\bissue{5}),
\bfpage{720}--\blpage{725}
(\byear{2023})
\doiurl{10.1038/s41567-023-01948-1}
\end{barticle}
\endbibitem

\bibitem[\protect\citeauthoryear{DeSalvo et~al.}{2019}]{DeSalvo2019}
\begin{barticle}
\bauthor{\bsnm{DeSalvo}, \binits{B.J.}},
\bauthor{\bsnm{Patel}, \binits{K.}},
\bauthor{\bsnm{Cai}, \binits{G.}},
\bauthor{\bsnm{Chin}, \binits{C.}}:
\batitle{{Observation of fermion-mediated interactions between bosonic atoms}}.
\bjtitle{Nature}
\bvolume{568}(\bissue{7750}),
\bfpage{61}--\blpage{64}
(\byear{2019})
\doiurl{10.1038/s41586-019-1055-0}
\end{barticle}
\endbibitem

\bibitem[\protect\citeauthoryear{Patel et~al.}{2023}]{Patel2023}
\begin{barticle}
\bauthor{\bsnm{Patel}, \binits{K.}},
\bauthor{\bsnm{Cai}, \binits{G.}},
\bauthor{\bsnm{Ando}, \binits{H.}},
\bauthor{\bsnm{Chin}, \binits{C.}}:
\batitle{{Sound Propagation in a Bose-Fermi Mixture: From Weak to Strong
  Interactions}}.
\bjtitle{Physical Review Letters}
\bvolume{131}(\bissue{8}),
\bfpage{083003}
(\byear{2023})
\doiurl{10.1103/PhysRevLett.131.083003}
\end{barticle}
\endbibitem

\bibitem[\protect\citeauthoryear{Yan et~al.}{2024}]{Yan2024a}
\begin{barticle}
\bauthor{\bsnm{Yan}, \binits{Z.Z.}},
\bauthor{\bsnm{Ni}, \binits{Y.}},
\bauthor{\bsnm{Chuang}, \binits{A.}},
\bauthor{\bsnm{Dolgirev}, \binits{P.E.}},
\bauthor{\bsnm{Seetharam}, \binits{K.}},
\bauthor{\bsnm{Demler}, \binits{E.}},
\bauthor{\bsnm{Robens}, \binits{C.}},
\bauthor{\bsnm{Zwierlein}, \binits{M.}}:
\batitle{{Collective flow of fermionic impurities immersed in a Bose–Einstein
  condensate}}.
\bjtitle{Nature Physics}
(\byear{2024})
\doiurl{10.1038/s41567-024-02541-w}
\end{barticle}
\endbibitem

\bibitem[\protect\citeauthoryear{Rakshit et~al.}{2019a}]{Rakshit2019a}
\begin{barticle}
\bauthor{\bsnm{Rakshit}, \binits{D.}},
\bauthor{\bsnm{Karpiuk}, \binits{T.}},
\bauthor{\bsnm{Brewczyk}, \binits{M.}},
\bauthor{\bsnm{Gajda}, \binits{M.}}:
\batitle{{Quantum Bose-Fermi droplets}}.
\bjtitle{SciPost Physics}
\bvolume{6}(\bissue{6}),
\bfpage{079}
(\byear{2019})
\doiurl{10.21468/SciPostPhys.6.6.079}
\end{barticle}
\endbibitem

\bibitem[\protect\citeauthoryear{Rakshit et~al.}{2019b}]{Rakshit2019}
\begin{barticle}
\bauthor{\bsnm{Rakshit}, \binits{D.}},
\bauthor{\bsnm{Karpiuk}, \binits{T.}},
\bauthor{\bsnm{Zin}, \binits{P.}},
\bauthor{\bsnm{Brewczyk}, \binits{M.}},
\bauthor{\bsnm{Lewenstein}, \binits{M.}},
\bauthor{\bsnm{Gajda}, \binits{M.}}:
\batitle{{Self-bound Bose–Fermi liquids in lower dimensions}}.
\bjtitle{New Journal of Physics}
\bvolume{21}(\bissue{7}),
\bfpage{073027}
(\byear{2019})
\doiurl{10.1088/1367-2630/ab2ce3}
\end{barticle}
\endbibitem

\end{thebibliography}

\end{document}